\documentclass[12pt]{article}
\usepackage{latexsym}
\catcode`\@=11

\@addtoreset{equation}{section}
\def\theequation{\arabic{section}.\arabic{equation}}
     
\catcode`\@=11
\def\thesection{\arabic{section}}

\def\appendix{\setcounter{section}{0}
        \def\thesection{Appendix.}
        \def\theequation{\Alph{section}.\arabic{equation}}}
\def\section{\@startsection{section}{1}{\z@}{3.5ex plus 1ex minus
   .2ex}{2.3ex plus .2ex}{\large\bf}}
     
\def\eqnarray{\let\@currentlabel=\theequation\refstepcounter{equation}
    \global\@eqnswtrue
    \global\@eqcnt\z@\tabskip\@centering\let\\=\@eqncr
    $$\halign to \displaywidth\bgroup\@eqnsel\hskip\@centering
      $\displaystyle\tabskip\z@{##}$&\global\@eqcnt\@ne 
       \hfil${{}##{}}$\hfil
      &\global\@eqcnt\tw@ $\displaystyle\tabskip\z@{##}$\hfil 
       \tabskip\@centering&\llap{##}\tabskip\z@\cr}
\def\lefteqn#1{\hbox to 4\arraycolsep{$\displaystyle #1$\hss}}
\long\def\@makefntext#1{\parindent 0cm\noindent
\hbox to 1em{\hss$^{\@thefnmark}$}#1}

\def\IR{{\hbox{{\rm I}\kern-.2em\hbox{\rm R}}}}
\def\IH{{\hbox{{\rm I}\kern-.2em\hbox{\rm H}}}}
\def\IC{{\ \hbox{{\rm I}\kern-.6em\hbox{\bf C}}}}
\def\IZ{{\hbox{{\rm Z}\kern-.4em\hbox{\rm Z}}}}

\newcommand{\beq}{\begin{equation}}
\newcommand{\eeq}{\end{equation}}
\def\slash#1{\setbox0=\hbox{$#1$}#1\hskip-\wd0\hbox to\wd0{\hss\sl/\/\hss}}

\begin{document}

%%%%%%%%%%%%%%%%%%%%%%%%%%%%%%%%%%%%%%%%%%%
%     C I T E . S T Y
%     compressed lists of numerical citations: [11-16]
%     see also OVERCITE.STY and DRFTCITE.STY
%
%     Copyright (C) 1989-1992 by Donald Arseneau
%     These macros may be freely transmitted, reproduced, or modified for
%     non-commercial purposes provided that this notice is left intact.
%
%
%  \@citen contains the code that parses the list of names, ignoring
%  spaces after commas, writes the aux file \citation, and formats the
%  number list.  \citen can be used by itself to give citation numbers
%  without the other formatting; e.g., "See also ref.~\citen{junk}."
%
\def\citen#1{%
\edef\@tempa{\@ignspaftercomma,#1, \@end, }% ignore spaces in parameter list
\edef\@tempa{\expandafter\@ignendcommas\@tempa\@end}%
\if@filesw \immediate \write \@auxout {\string \citation {\@tempa}}\fi
\@tempcntb\m@ne \let\@h@ld\relax \let\@citea\@empty
\@for \@citeb:=\@tempa\do {\@cmpresscites}%
\@h@ld}
%
% for ignoring spaces in the input:
\def\@ignspaftercomma#1, {\ifx\@end#1\@empty\else
   #1,\expandafter\@ignspaftercomma\fi}
\def\@ignendcommas,#1,\@end{#1}
%
% For each citation, check if it is defined, if it is a number, and
% if it is a consecutive number that can be represented like 3-7.
%
\def\@cmpresscites{%
 \expandafter\let \expandafter\@B@citeB \csname b@\@citeb \endcsname
 \ifx\@B@citeB\relax % undefined
    \@h@ld\@citea\@tempcntb\m@ne{\bf ?}%
    \@warning {Citation `\@citeb ' on page \thepage \space undefined}%
 \else%  defined
    \@tempcnta\@tempcntb \advance\@tempcnta\@ne
    \setbox\z@\hbox\bgroup % check if citation is a number:
    \ifnum\z@<0\@B@citeB \relax
       \egroup \@tempcntb\@B@citeB \relax
       \else \egroup \@tempcntb\m@ne \fi
    \ifnum\@tempcnta=\@tempcntb % Number follows previous--hold on to it
       \ifx\@h@ld\relax % first pair of successives
          \edef \@h@ld{\@citea\@B@citeB}%
       \else % compressible list of successives
%         % use \hbox to avoid easy \exhyphenpenalty breaks
          \edef\@h@ld{\hbox{--}\penalty\@highpenalty \@B@citeB}%
       \fi
    \else   %  non-successor--dump what's held and do this one
       \@h@ld \@citea \@B@citeB \let\@h@ld\relax
 \fi\fi%
 \let\@citea\@citepunct
}
%
%%    To put space after the comma, use:
\def\@citepunct{,\penalty\@highpenalty\hskip.13em plus.1em minus.1em}%
%%    For no space after comma, use:
%% \def\@citepunct{,\penalty\@highpenalty}%
%%
%
%  Make \@citex refer to \citen:
%
\def\@citex[#1]#2{\@cite{\citen{#2}}{#1}}%
%
%  Replacement for \@cite.  Give one normal space before the citation,
%  set high penalties for linebreaks,
%
\def\@cite#1#2{\leavevmode\unskip
  \ifnum\lastpenalty=\z@ \penalty\@highpenalty \fi % highpenalty before
  \ [{\multiply\@highpenalty 3 #1% % triple-highpenalties within list
      \if@tempswa,\penalty\@highpenalty\ #2\fi % and before note.
    }]\spacefactor\@m}
\let\nocitecount\relax  % in case \nocitecount was used for drftcite
%
%%%%%%%%%%%%%%%%%%%%%%%%%%%%%%%%%%%%%%%%%%%% 
\begin{titlepage}
\vspace{.5in}
\begin{flushright}
UCD-2001-04\\
gr-qc/0108040\\
March 2001\\
\end{flushright}
\vspace{.5in}
\begin{center}
{\Large\bf
 Quantum Gravity: a Progress Report}\\
\vspace{.4in}
{S.~C{\sc arlip}\footnote{\it email: carlip@dirac.ucdavis.edu}\\
       {\small\it Department of Physics}\\
       {\small\it University of California}\\
       {\small\it Davis, CA 95616}\\{\small\it USA}}
\end{center}

\vspace{.5in}
\begin{center}
{\large\bf Abstract}
\end{center}
\begin{center}
\begin{minipage}{4.75in}
{\small The problem of reconciling general relativity and quantum theory
has fascinated and bedeviled physicists for more than 70 years.  Despite
recent progress in string theory and loop quantum gravity, a complete
solution remains out of reach.  I review the status of the continuing effort 
to quantize gravity,  emphasizing the underlying conceptual issues and the 
various attempts to come to grips with them.
}
\end{minipage}
\end{center}
\end{titlepage}
\addtocounter{footnote}{-1}

\section{Introduction}

The two main pillars of modern physics, it is often said, are general 
relativity and quantum theory.  Phenomena at large scales are governed
by gravitational interactions, and observations from cosmological 
distances to millimeter scales \cite{EotWash} are well-described by 
general relativity.  Phenomena at small scales are dominated by strong 
and electroweak interactions, and observations at distances ranging 
from a fraction of a millimeter \cite{superpos} down to $10^{-19}$ 
meters are well-described by quantum mechanics and quantum field 
theory.  No known fundamental interaction falls outside this framework.

When we look more closely, however, it is not so clear that these
two pillars are part of the same edifice.  The foundations of general 
relativity---a dynamical spacetime, with no preferred reference 
frame---clash with the needs of quantum theory, which in its standard 
formulations requires a fixed background and a preferred splitting of 
spacetime into space and time.  Despite some 70 years of active research, 
no one has yet formulated a consistent and complete quantum theory of 
gravity.

The failure to quantize gravity rests in part on technical difficulties.
General relativity is, after all, a complicated and highly nonlinear theory.  
Indeed, it was not until 1986 that it was finally shown that conventional 
quantum field theoretic techniques fail \cite{Goroff}.  But the real problems 
are almost certainly  deeper: quantum gravity requires a quantization of 
spacetime itself, and at a fundamental level we do not know what that means.

The two leading candidates for a quantum theory of gravity  today
are string theory and loop quantum gravity.  I will not try to 
review these approaches in detail; readers will find a thorough 
introduction to string theory in Polchinski's textbook \cite{Polchinski}, 
and an excellent review of quantum geometry in Rovelli's article \cite{Rovelli}.  
I will instead try to give a broader overview, concentrating on underlying 
conceptual problems and the attempts to resolve them. 

While string theory and loop quantum gravity have many attractive features,
there is not, at this writing, a compelling argument that either is the correct
quantum theory of gravity.  Nor is there much observational evidence to
point us in any particular direction.  Quantum gravity remains a theorists' 
playground, an arena for ``theoretical experiments,'' some of them quite
adventurous, which may or may not stand the test of time.  Some of the ideas 
I discuss here will survive, but others will undoubtedly be mere historical 
curiosities a decade from now.   

This paper should be read as an outline and a guide to further reading.
I make no claims of being complete or unbiased.  For complementary reading 
and some different perspectives, I suggest Rovelli's recent summary of the 
history of quantum gravity \cite{Rovelli2}, Isham's 1991 review \cite{Isham0},  
and Au's interviews with several leading practitioners \cite{Au}.  Readers
may also want to look at the new collection {\it Physics meets philosophy 
at the Planck scale\/} \cite{Calendar}.

Readers should be warned that conventions on indices, units, signs,
and the like vary widely from paper to paper.  Equations here should not 
be applied elsewhere without carefully checking conventions.  Finally, let 
me stress that my references are by no means comprehensive.  In particular, 
while I have tried to avoid major historical inaccuracies, I often cite later 
reviews rather than original papers.  I apologize to the people whose work 
I have neglected, misunderstood, or mutilated in an effort to keep this review 
finite in length.

\section{Why quantum gravity?}\setcounter{footnote}{0}

Before undertaking such a difficult task as the quantization of general
relativity, one should first ask, ``Is this really necessary?''  The problems
addressed by general relativity and those addressed by quantum theory
typically arise at very different length and energy scales, and there is not 
yet any direct experimental evidence that gravity is quantized.  Perhaps 
general relativity is fundamentally different; maybe we don't need to 
quantize it.

The stock reply is an appeal to the unity of physics.  The historical trend of 
fundamental physics has certainly been toward unity, from Maxwell's unification 
of electricity and magnetism to the Weinberg-Salam electroweak model and 
the ubiquity of gauge theories.  But such a historical argument is not entirely 
convincing; let us try to go further.

Consider a scheme in which the gravitational field is not quantized.  The obvious 
objection is that such a theory could lead to a violation of the uncertainty principle: 
gravity could be used to simultaneously determine the position and momentum 
of a particle to arbitrary accuracy.  This problem has been studied by Eppley and 
Hannah \cite{Eppley}, who show that if ``measurement'' by a gravitational wave 
leads to wave function collapse, the uncertainty relations can be saved only by 
sacrificing conservation of momentum.  If, on the other hand, gravitational 
``measurements'' do not cause wave function collapse, then gravitational interactions 
with quantum matter could be used to transmit an observable signal faster than light.  
One might instead appeal to the Everett (``many worlds'') interpretation of quantum 
theory, but  unquantized gravity in this picture is physically unrealistic and, in fact,
experimentally excluded \cite{Page}: if gravity were not quantized, a quantum 
superposition of macroscopically separated distributions of matter would produce 
a gravitational field pointing toward the ``average'' center of mass rather than the 
observed definite position.

Further difficulties arise in more detailed proposals.  The simplest coupling of 
quantum theory and classical gravity, often called ``semiclassical gravity,'' was 
proposed by M{\o}ller \cite{Moller} and Rosenfeld \cite{Rosenfeld}.  In this 
approach, the Einstein field equations take the form
\beq
G_{\mu\nu} = 8\pi G \langle\psi| T_{\mu\nu}|\psi\rangle ,
\label{a1}
\eeq
where the operator-valued stress-energy tensor of matter is replaced by an 
expectation value.  Note that some change of this sort is necessary: if matter is 
quantized, the stress-energy tensor is an operator, and cannot be simply set 
equal to the $c$-number Einstein tensor.

To investigate such a model, it is useful to start with a simpler model of Newtonian 
gravity coupled to nonrelativistic quantum matter via the Schr{\"o}dinger equation 
\cite{Diosi}.  For a particle of mass $m$, we have
\beq
i\hbar{\partial\psi\over\partial t} = -{\hbar^2\over2m}\nabla^2\psi
   + mV\psi , \quad \hbox{with}\quad
\nabla^2V = 4\pi G m|\psi|^2
\label{a2}
\eeq
where the second equality is the Poisson equation for the Newtonian 
gravitational potential of a mass distribution $m|\psi|^2$.  As an 
approximation of an exact theory, these equations are the analog of the 
Hartree approximation, and present no problem.  If they are taken to be 
fundamental, though, they clearly lead to a nonlinear Schr{\"o}dinger 
equation; the principle of superposition fails, and with it go the foundations
of conventional quantum mechanics.  The full general relativistic version of 
this nonlinearity has been discussed by Kibble and Randjbar-Daemi \cite{Kibble}.  
It is not certain that the resulting theory is inconsistent with existing experiment 
\cite{Bertolami}, but the deep incompatibility with standard quantum theory 
is clear.  

While there are continuing efforts to understand how to couple a quantum 
system to a classical one \cite{Halliwell}, these arguments strongly suggest 
that an internally consistent model of the physical world requires that either 
general relativity or quantum mechanics be changed.  The usual choice is to 
demand the quantization of general relativity, although some, notably Penrose 
\cite{Penrose}, argue that perhaps it is quantum mechanics that ought to be 
modified.  I suspect the preference for the former comes in part from ``majority 
rule''---most interactions are very successfully described by quantum field theory, 
with general relativity standing alone outside the quantum framework---and 
in part from the fact that we already know a good deal about how to quantize a 
classical theory, but almost nothing about how to consistently change quantum 
theory.

There is a second line of argument for quantizing gravity, arising more from
hope than necessity.  Both classical general relativity and quantum field theory
have serious limitations, and there is some reason to believe that quantum
gravity may offer a cure.

Wheeler's famous characterization \cite{MTW} of gravitational collapse as 
``the greatest crisis in physics of all time'' may be hyperbole, but general
relativity's prediction of the inevitability of singularities is certainly a cause
for concern.  Cosmology faces a similar problem: an initial singularity not
only fails to provide an adequate set of initial conditions, it even removes the
point at which initial conditions could be imposed.  While no one has proven
that the quantization of gravity will eliminate singularities, this is the sort of 
thing one might expect from a quantum theory.  A proper treatment of quantum 
gravity might even determine initial conditions for the Universe, making 
cosmology a completely predictive science \cite{Hartle}.

Quantum field theory, in turn, has its own problems, in the form of the infinities
that plague perturbation theory.  From the modern point of view \cite{Weinberg}, 
most quantum field theories are really ``effective field theories,'' in which the 
divergences reflect our ignorance of physics at very high energies.  It has long 
been speculated that the missing ingredient is quantum gravity, which has a 
natural length scale and might provide an automatic cutoff at the Planck energy 
\cite{Landau,Pauli,Deser}.  While there is no proof that such a cutoff occurs, 
there are some suggestive results: for example, when gravity is included, certain 
infinite sets of divergent Feynman diagrams can be resummed to give finite results 
\cite{DeWitt,Khriplovich,Isham2}.

We thus approach quantum gravity with a mixture of hope and fear: hope that
it can solve some fundamental problems in general relativity and quantum field
theory and perhaps offer us a unified picture of physics, and fear that a failure
will demonstrate underlying flaws in the physics we think we now understand.
Given these powerful motivations, we turn to our next question: Why has
general relativity not yet been quantized?

\section{Why is gravity not yet quantized? \label{secb}}\setcounter{footnote}{0}

The first technical papers on quantum gravity were written by Rosenfeld in 
1930 \cite{Rosenfeld2}.  The list of researchers who have worked on the problem 
since then reads like a {\em Who's Who\/} of modern physics, including  
ten Nobel Laureates---Einstein, Bohr, Heisenberg, Dirac, Pauli, Schwinger, 
Feynman, Veltman, 't Hooft, and Weinberg.  In one sense, the work has been 
very successful, leading to much that we now take for granted: gauge fixing and 
Faddeev-Popov ghosts, the background field method and the effective 
action, and much of what we know about constrained dynamics.  What it has 
{\em not\/} led to is a quantum theory of gravity.

Quantum gravity is undoubtedly technically hard, but this failure has deeper 
roots in our lack of understanding of what ``quantized spacetime'' might mean.  
This is a bit of a clich{\'e}, and deserves further explanation:  

In an ordinary field theory on a fixed background spacetime $M$, the points in 
$M$ are physically meaningful.  It makes sense, for example, to speak of ``the 
value of the electromagnetic field at the point $x$.''  General relativity, in 
contrast, is invariant under diffeomorphisms, ``active'' coordinate transformations 
that move points in $M$, and points no longer have any independent meaning.
Consider, for instance, a small empty region (a ``hole'') $V\subset M$.  Let
$f: M\rightarrow M$ be a diffeomorphism that reduces to the identity outside 
$V$ but differs from the identity inside $V$.  By assumption, $f$ does not affect 
matter---there is no matter in $V$---but merely ``moves points'' in empty 
spacetime.  In a noncovariant theory, or more generally a theory with a fixed 
background structure, $M$ and $f(M)$ are distinct manifolds, and one can talk
about ``a point $x\in V$.''  In general relativity, though---at least in the standard 
physicists' interpretation---$M$ and $f(M)$ are {\em identical\/}, even though 
we have ``moved some points,''  and there is no way to distinguish $x$ and $f(x)$.

This is a short version of Einstein's ``hole argument,''  and was one of his reasons 
for initially rejecting general covariance \cite{Stachel,Norton}.  For us, the 
significance is that we cannot view the metric as merely a superstructure sitting atop 
a physically meaningful set of points that make up a spacetime.  The manifold and 
the geometry are fundamentally inseparable, and quantizing the geometry really 
does mean quantizing the spacetime itself.

This problem appears in a number of guises:

\begin{enumerate}
\item{\bf General covariance vs.\ locality:}  The fundamental symmetry of 
general relativity is general covariance (strictly speaking, diffeomorphism 
invariance), the lack of dependence of physical quantities on the choice of 
coordinates.  Observables in quantum gravity should presumably respect this 
symmetry \cite{Bergmann0}.  But diffeomorphism-invariant observables in 
general relativity are necessarily nonlocal \cite{Torre}, essentially because 
active coordinate transformations ``move points'' and cannot preserve a 
quantity defined by its value at individual points.   
\item{\bf The ``problem(s) of time'':}  Time plays two vital roles in quantum 
theory: it determines the choice of canonical positions and momenta, and it fixes 
the normalization of the wave function, which must be normalized to one {\em at 
a fixed time} \cite{Unruh}.  In general relativity, though, there is no preferred 
``time slicing'' of spacetime into spatial hypersurfaces.  This has many consequences, 
discussed in detail in review papers by Kucha{\v r} \cite{Kuchar} and Isham 
\cite{Isham2b}; I will only touch on a few highlights.
\begin{itemize}
\item The natural Hamiltonian in general relativity is a constraint,
which, up to possible boundary terms, is identically zero for physical states.  In
retrospect, this is not surprising: a time translation $t\rightarrow t+\delta t$ can
be viewed as a coordinate transformation, and general relativity is invariant under
such transformations \cite{Misner,Bergmann}.  Similarly, if we impose the
natural requirement that observables commute with constraints, then all
observables must be constants of motion.
\item Quantum field theory includes causality as a fundamental axiom: fields at 
points separated by spacelike intervals must commute.  But if the metric itself is 
subject to quantum fluctuations, we can no longer tell whether the separation 
between two points is spacelike, null, or timelike.  Quantum fluctuations 
of the metric can exchange past and future.  This has led to speculations that 
causality requirements might compel drastic changes in the starting point of 
quantization \cite{Penrose2,Sorkin}.
\item In classical general relativity, the evolution of a configuration from an 
initial to a final spatial hypersurface is independent of the choice of time 
coordinate in the intervening spacetime. While this may still be true in quantum 
gravity with a proper choice of operator ordering \cite{Cosgrove}, the issue is 
far from being settled.  Indeed, for even as simple a system as a scalar field in 
a flat spacetime, different choices of intermediate time slicing can lead to 
inequivalent quantum evolution \cite{Varadarajan}.
\item The obvious candidates for wave functions in quantum gravity, solutions
of the Wheeler-DeWitt equation, are not normalizable.  This is to be expected:
because of general covariance, time enters into the wave function only 
implicitly through the metric \cite{Baierlein}, and the normalization secretly 
involves an integral over time.  It may be possible to cure this problem by 
``gauge-fixing the inner product'' \cite{Woodard}, but solutions of this sort are, 
at best, technically very difficult \cite{Marolf,Giulini}.
\end{itemize}
It is tempting to sidestep these problems by defining time as ``the reading of a 
clock.''  But a clock only measures time along its world line; to define time globally, 
one needs a space-filling ``reference fluid,'' which then has a back reaction on the 
gravitational field \cite{Kuchar}.  Worse, a clock made of quantum matter cannot 
be reliable: any clock built from matter with a positive Hamiltonian has a finite 
probability of sometimes running backward \cite{Wald3}, and thus cannot be used 
to consistently normalize wave functions.
\item{\bf The reconstruction problem:}  We saw above that observables in quantum
gravity must be nonlocal constants of motion.  Even if we find a set of observables,
we must still figure out how to reconstruct the standard local description as the 
classical limit.  This problem is already present in classical general relativity: to 
accurately test Solar System predictions, for instance, one must replace 
coordinate-dependent quantities (``the position of the Moon'') with invariants 
(``the round trip time for a radar pulse reflected by the Moon, as measured 
by a particular clock'').  In quantum gravity, it is harder (how does one specify 
``the Moon'' or ``this clock''?);  even for the simple model of general relativity in 
2+1 dimensions, such a reconstruction requires a complete understanding of the 
space of solutions of the classical field equations \cite{Carlip,Carlip2}.
\item{\bf Small-scale structure:} Computations in standard quantum field theory 
are almost always perturbative, involving expansions around a simple vacuum
state.  When the vacuum is not simple---in low energy quantum chromodynamics, 
for instance---these methods often fail.  When general relativity is treated as an 
ordinary quantum field theory, the usual choice for the vacuum is flat Minkowski 
space.  This is a reasonable guess, given the positive energy theorem, which states
that Minkowski space is the lowest energy asymptotically flat solution of the classical 
Einstein field equations.  

But it is not clear that the ground state of quantum gravity can be described as a 
classical smooth manifold at all.  Indeed, various analyses of the measurement 
process suggest that quantum gravity has a minimum length scale, below which a 
classical description makes no sense \cite{Garay}.  The analyses differ in detail, 
and should not be taken as the final word, but it is not unreasonable to expect a 
modified uncertainty relationship of the form 
\beq
\Delta x \ge {\hbar\over\Delta p} + L^2 {\Delta p\over\hbar} ,
\label{b1}
\eeq
where $L$ is of the order of the Planck length, $L_{\hbox{\scriptsize Planck}} =  
({\hbar G/c^3})^{1/2}   \approx 10^{-35}\ \mathrm{m.}$  If this is the case, 
perturbation theory around a smooth classical background may simply not make 
sense.  The picture is further complicated by Wheeler's suggestion \cite{Wheeler} 
that even the topology of spacetime may be subject to quantum fluctuations,
leading to microscopic ``spacetime foam.''
\item{\bf Large-scale structure and scattering states:} It is very difficult to
describe exact states in an interacting quantum field theory.  We usually avoid 
this problem by focusing on $S$-matrix elements between asymptotic states.  
As a practical matter, this works as long as states far from the interaction region 
are nearly those of a free field theory.  But in general relativity, it is doubtful that 
free field theory is a good approximation even asymptotically, since quantum 
fluctuations of the short-distance structure of spacetime will still be present 
\cite{Tsamis}.
\item{\bf The ``wave function of the Universe'':} A key motivation for quantum 
gravity is the need to understand quantum cosmology, the quantum mechanics 
of the Universe as a whole.  But the observer is a part of the Universe, and one can 
no longer make the conventional split between observer and observed.  We must 
thus face a whole set of questions about the meaning of quantum mechanics that 
can usually be ignored \cite{Bell}: When do wave functions collapse?  What does 
it mean to assign a probability to a unique system?  What makes an ``observer'' 
special in quantum theory? 
\end{enumerate}

Given these fundamental issues, it is perhaps remarkable that any progress at all 
has been made in quantizing gravity.  But despite the difficulty of the problem, 
a good deal has been learned.  Existing approaches to quantum gravity fall into 
two broad categories, ``covariant'' and ``canonical.''  Covariant quantization treats 
diffeomorphism invariance as fundamental, and tries to manifestly preserve this 
symmetry.  This usually requires perturbative quantization around a fixed background.  
Canonical quantization treats the symplectic structure of quantum mechanics as 
fundamental, and splits the classical variables into ``positions'' and ``momenta'' 
from the start.  This allows a nonperturbative treatment, but usually at the 
expense of manifest covariance.  There is a long history of debate between 
advocates of these two philosophies, which has mainly served to clarify the
weaknesses of each.  Given that neither approach can yet boast of any 
overwhelming success, I will not attempt to choose between them, but will 
review both.

\section{Classical preliminaries}\setcounter{footnote}{0}

The real world is, presumably, quantum mechanical, and ideally we should 
start with a quantum theory and obtain classical physics as a limiting 
case.  In practice, though, we rarely know how to formulate a quantum theory 
directly from first principles; the best we can do is to start with a classical theory 
and ``quantize'' it.  We must therefore  understand something of classical general 
relativity before discussing quantum gravity.   General relativity can be described 
in a number of different ways, each of which suggests a different approach to 
quantization.  I shall now briefly review these starting points.\footnote{Although
generalizations are not hard, I shall usually restrict myself to spacetimes of the 
observed four dimensions.  My conventions are as follows: Greek letters $\lambda$, 
$\mu$, \dots are spacetime coordinate indices, ranging from $0$ to $3$; lower case 
Roman letters $i$, $j$, \dots are spatial indices at a fixed time, ranging from $1$ 
to $3$; capital Roman letters $I$, $J$, \dots are ``tangent space indices,'' ranging 
from $0$ to $3$, that label vectors in an orthonormal tetrad, and are subject to 
local Lorentz ($\hbox{SO}(3,1)$) transformations; hatted capital Roman letters 
${\hat I}$, ${\hat J}$,  \dots are gauge-fixed tangent space indices, ranging from 
$1$ to $3$ and subject to local $\hbox{SO}(3)$ transformations.}

\subsection{Covariant formalism: second order form \label{secca}}

We begin with the Einstein-Hilbert action on a manifold $M$, 
\beq
I = {1\over16\pi G}\int_M d^4x \sqrt{-g}(R - 2\Lambda) 
 \pm {1\over8\pi G}\int_{\partial M} d^3x \sqrt{q}\,K
 + I_{\hbox{\scriptsize matter}} ,
\label{c1}
\eeq
where $q_{ij}$ is a (fixed) induced three-metric at the boundary of $M$ and $K$ is 
the mean extrinsic curvature of the boundary (see section \ref{seccb}).  The 
boundary term in (\ref{c1}) may be unfamiliar; it is needed to cancel terms in the 
variation of the action that arise from integration by parts \cite{Gibbons3,Yorkx}, 
and appears with a positive sign  for spacelike components of $\partial M$ and a 
negative sign for timelike components.  

The  action (\ref{c1}) is invariant under diffeomorphisms mapping $M$ to $M$.  
For an infinitesimal diffeomorphism $x^\mu\rightarrow x^\mu-\xi^\mu$, the fields 
transform as
\beq
\delta g_{\mu\nu} 
  = {\cal L}_\xi g_{\mu\nu} = \nabla_\mu\xi_\nu + \nabla_\nu\xi_\mu ,\qquad
\delta\varphi = {\cal L}_\xi\varphi ,
\label{c5}
\eeq
where ${\cal L}_\xi$ denotes the Lie derivative and $\varphi$ represents an arbitrary 
collection of matter fields.  Diffeomorphisms can be viewed as ``active'' coordinate 
transformations, and diffeomorphism invariance is essentially general covariance.  
But as noted earlier, the active view of these transformations has a deeper philosophical 
import: it highlights the fact that spacetime points have no independent physical meaning.

Although the action (\ref{c1}) is usually introduced geometrically, there is an 
interesting alternative.  Start with a flat spacetime with metric $\eta_{\mu\nu}$, 
and postulate a massless spin two tensor field $h_{\mu\nu}$ that couples universally 
to the stress-energy tensor.  As a first approximation, we can write a field equation 
for ${\bar h}_{\mu\nu} = h_{\mu\nu} - {1\over2}\eta_{\mu\nu}\eta^{\rho\sigma}
h_ {\rho\sigma}$ in the form
\beq
\Box{\bar h}_{\mu\nu} 
  - \eta^{\sigma\tau}(\partial_\mu\partial_\sigma{\bar h}_{\nu\tau}
  + \partial_\nu\partial_\sigma{\bar h}_{\mu\tau}) 
  + \eta_{\mu\nu}\eta^{\pi\rho}\eta^{\sigma\tau}\partial_\pi\partial_\sigma
  {\bar h}_{\rho\tau}  = 16\pi G T_{\mu\nu} , 
\label{c6}
\eeq
where the coefficients are fixed by energy conservation and the requirement that 
$h_{\mu\nu}$ be pure spin two \cite{Feynman}.  These equations can be obtained 
from a Lagrangian ${\cal L}^{(2)}$ quadratic in $h_{\mu\nu}$.  But (\ref{c6}) 
is only a ``first approximation'': the right-hand side should include the stress-energy 
tensor of $h_{\mu\nu}$ itself, which is quadratic in $h_{\mu\nu}$.  To obtain such
a source term from our Lagrangian, we must include a cubic term ${\cal L}^{(3)}$,
which in turn leads to a cubic term to the stress-energy tensor, requiring a quartic 
term ${\cal L}^{(4)}$, etc.  With a clever choice of variables, the resulting series can 
be made to terminate, and the sum leads almost uniquely to the Einstein-Hilbert 
action for the metric $g_{\mu\nu} = \eta_{\mu\nu} + h_{\mu\nu}$ \cite{Deser0}.   
A similar derivation starts with the quantum field theory of a massless spin two field, 
and shows that the low-energy limit is necessarily Einstein gravity \cite{Boulware}.  
It was this property that first led to the realization that string theory---then regarded 
as a theory of hadrons---might be connected with gravity \cite{Scherk}.

\subsection{Canonical formalism: second order form \label{seccb}}

\begin{figure}
\begin{picture}(200,130)(-90,-40)

\qbezier(-0,110)(40,85)(42,50)         % top
\qbezier(170,110)(210,85)(212,50)
\put(0,110){\line(1,0){170}}
\put(42,50){\line(1,0){170}}
\put(202,86){$\Sigma_{t+dt}$}

\qbezier(10,20)(50,-5)(52,-40)        % bottom
\qbezier(180,20)(220,-5)(222,-40)
\put(10,20){\line(1,0){170}}
\put(52,-40){\line(1,0){170}}
\put(214,-8){$\Sigma_t$}

\put(100,-10){\circle*{2}}             % x on bottom
\put(103,-13){$x^i$}

\put(100,-10){\line(0,1){60}}          % Ndt
%\put(100,43){\circle*{1}}
%\put(100,47){\circle*{1}}
%\put(100,51){\circle*{1}}
\put(100,55){\circle*{1}}
\put(100,59){\circle*{1}}
\put(100,63){\circle*{1}}
\put(100,67){\circle*{1}}
\put(103,26){$Ndt$}

\put(100,71){\circle*{2}}              % x-Ndt on top
\put(103,70){$x^i-N^idt$}

\put(100,-10){\line(-1,3){20}}         % ds
%\put(82,44){\circle*{1}}
%\put(80,50){\circle*{1}}
\put(78,56){\circle*{1}}
\put(76,62){\circle*{1}}
\put(74,68){\circle*{1}}
\put(72,74){\circle*{1}}
\put(70,80){\circle*{1}}
\put(68,86){\circle*{1}}
\put(73,26){$ds$}

\put(66,92){\circle*{2}}              % x+dx on top
\put(66,94){$x^i+dx^i$}

\qbezier(100,71)(88,88)(66,92)
\end{picture}
\caption{The ADM decomposition expresses the Lorentzian version of
the Pythagoras theorem. \label{fig1}}
\end{figure}
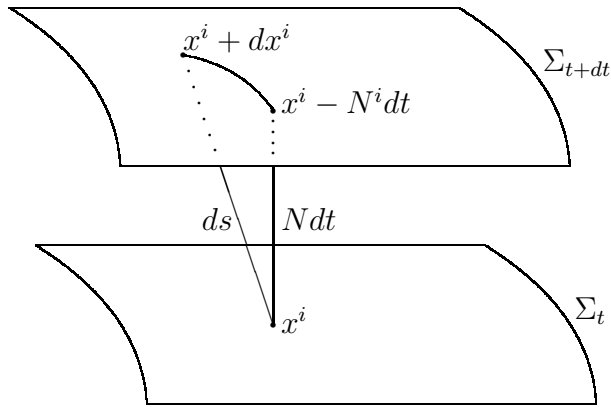

We next turn to the Hamiltonian, or canonical, formulation of general relativity.
Such a formulation requires a choice of time, that is, a slicing of spacetime into 
preferred spatial hypersurfaces.  The Arnowitt-Deser-Misner (ADM) approach  
\cite{ADM} starts with a slicing of $M$ into constant-time hypersurfaces 
$\Sigma_t$, each equipped with coordinates $\{x^i\}$ and a three-metric $q_{ij}$ 
with determinant $q$ and inverse $q^{ij}$. (For equivalent ``modern'' descriptions 
in terms of four-vectors, see  \cite{Isham2b} or Appendix E of \cite{Wald}.)  To 
obtain the four-geometry, we start at a point on $\Sigma_t$ with coordinates $x^i$, 
and displace it infinitesimally normal to $\Sigma_t$.  The resulting  change in proper 
time can be written as $d\tau = Ndt$, where $N$ is called the lapse function.  In a 
generic coordinate system, though, such a displacement will also shift the spatial 
coordinates: $x^i(t+dt) = x^i(t)-N^idt$, where $N^i$ is called the shift vector.  By 
the Lorentzian version of the Pythagoras theorem  (see figure~\ref{fig1}), the interval 
between $(t,x^i)$ and $(t+dt,x^i+dx^i)$ is then
\beq
ds^2 = -N^2dt^2 + q_{ij}(dx^i+N^idt)(dx^j+N^jdt) .
\label{c9}
\end{equation}

It is customary in the canonical formalism to establish new conventions that 
emphasize the role of the hypersurface $\Sigma$.  Spatial indices will now be 
lowered and raised with the spatial metric $q_{ij}$ and its inverse $q^{ij}$,
and not with the full spacetime metric.  (Note that $q^{ij}$ is not simply the 
spatial part of the four-metric $g^{\mu\nu}$.)  When there is a likelihood of 
confusion between three- and four-dimensional objects, I will use a prefix 
${}^{(3)}$ or ${}^{(4)}$.

Expressed in terms of the ADM metric, the Einstein-Hilbert action is a function 
of $q_{ij}$ and its first time derivative, or equivalently $q_{ij}$ and the extrinsic 
curvature $K_{ij}$ of the time slice $\Sigma_t$ viewed as an embedded hypersurface.  
A straightforward computation \cite{MTW} shows that the canonical momentum 
conjugate to $q_{ij}$ is 
\beq
\pi^{ij} = {\partial{\cal L}\over\partial(\partial_tq_{ij})} = {1\over16\pi G}
  \sqrt{q}(K^{ij} - q^{ij}K) ,
\label{c13}
\eeq
where $K = q^{ij}K_{ij}$ (sometimes written $\mathrm{Tr}K$) is the mean 
extrinsic curvature.  The action (\ref{c1})
becomes
\beq
I = \int dt\int_\Sigma d^3x\,\left(
  \pi^{ij}\partial_tq_{ij} - N{\cal H} - N_i{\cal H}^i\right) 
  + \hbox{\em boundary terms},
\label{c15}
\eeq
where
\beq
{\cal H} = 16\pi G{1\over \sqrt{q}}(\pi_{ij}\pi^{ij} - {1\over2}\pi^2)
  - {1\over16\pi G}\sqrt{q}\,({}^{(3)}\!R - 2\Lambda)
\label{c16}
\eeq
is known as the Hamiltonian constraint and
\beq
{\cal H}^i = -2 {}^{(3)}\nabla_j\pi^{ij}
\label{c17}
\eeq
are the momentum constraints.  The field equations are the standard Hamilton's
equations of motion for this action, with the Poisson brackets
\beq
\{q_{ij}(x), \pi^{kl}(x')\} = {1\over2}(\delta_i^k\delta_j^l
  + \delta_i^l\delta_j^k)\tilde\delta^3(x-x')  ,
\label{c14}
\eeq
where $\tilde\delta^3(x-x')$ is the metric-independent (``densitized'') delta 
function.   Note that the Hamiltonian constraint is quadratic in the momenta 
$\pi^{ij}$  and nonpolynomial in the positions $q_{ij}$.  Both of these features 
will return to plague us when we try to quantize this system.

The action (\ref{c15}) contains no time derivatives of the lapse and shift functions; 
$N$ and $N_i$ are not dynamical variables, but only Lagrange multipliers.  Their 
variation leads to the field equations ${\cal H}=0$ and ${\cal H}^i=0$, implying 
that the ``Hamiltonian term'' in the action (\ref{c15}) vanishes on shell, up to 
possible boundary terms.  The conditions ${\cal H}=0$ and ${\cal H}^i=0$ 
themselves involve no time derivatives, and are thus not ordinary dynamical 
equations of motion.  Rather, they are constraints on the possible initial values 
of the fields $q_{ij}$ and $\pi^{ij}$.  In Dirac's terminology \cite{Dirac}, they 
are first class constraints: that is, the Poisson brackets of any pair of constraints 
is itself proportional to the constraints.

In any constrained Hamiltonian system, the first class constraints generate gauge 
transformations \cite{Henneaux}.   This statement is nontrivial, but plausible:
since the constraints vanish on physical states, their brackets with the canonical
variables should be ``zero,'' i.e., physically unobservable.  Since the fundamental 
symmetry of general relativity is diffeomorphism invariance, one might expect 
${\cal H}$ and ${\cal H}^i$ to generate diffeomorphisms.  This is almost the 
case---the constraints generate ``surface deformations,'' equivalent to 
diffeomorphisms when the field equations hold.  More specifically, if $\xi^\mu =
(0,\xi^i)$ is an infinitesimal deformation of the time slice $\Sigma_t$, the
transformation (\ref{c5}) is an ordinary canonical transformation, generated
by the momentum constraints ${\cal H}_i$.  If $\xi^0\ne0$, though, the
diffeomorphism $x^\mu\rightarrow x^\mu-\xi^\mu$ includes a time translation
from $\Sigma_t$ to $\Sigma_{t+\xi^0}$, and is in some sense ``dynamical.''
Such a diffeomorphism is generated by the Hamiltonian constraint $\cal H$,
but only on shell, that is, up to terms proportional to the equations of motion.

The entanglement of symmetry and dynamics is an instance of the ``problem of 
time'' discussed in section \ref{secb}.  In reappears in the Poisson algebra of the 
constraints, which is not the standard algebra of spacetime diffeomorphisms.  
Rather, the constraints give a representation of the surface deformation algebra 
\cite{Teitelboim,Brown}
\begin{eqnarray}
\{ {\hat\xi}_1, {\hat\xi}_2 \}_{\hbox{\scriptsize SD}}^\perp &=&
{\hat\xi}_1^i\partial_i{\hat\xi}_2^\perp - {\hat\xi}_2^i\partial_i{\hat\xi}_1^\perp
\nonumber\\
\{ {\hat\xi}_1, {\hat\xi}_2 \}_{\hbox{\scriptsize SD}}^i &=&
{\hat\xi}_1^j\partial_j{\hat\xi}_2^i - {\hat\xi}_2^j\partial_j{\hat\xi}_1^i
+ q^{ij}\left( {\hat\xi}_1^\perp\partial_j{\hat\xi}_2^\perp -
{\hat\xi}_2^\perp\partial_j{\hat\xi}_1^\perp \right) ,
\label{c23}
\end{eqnarray}
where ${\hat\xi}^\perp = N\xi^0$ and ${\hat\xi}^i = \xi^i + N^i\xi^0$.  The explicit 
presence of the metric on the right-hand side of (\ref{c23}) means that the algebra
of constraints is not a Lie algebra---it has ``structure functions'' rather than structure 
constants.  This considerably complicates quantization \cite{DeWitt2}.  

It has recently been observed that certain combinations of the constraints form a 
genuine Lie algebra \cite{Brown2,Kuchar2,Markopoulou,Kouletsis}, although still
not the algebra of spacetime diffeomorphisms.  For instance, if $G = {\cal H}^2 - 
q^{ij}{\cal H}_i{\cal H}_j$, the constraints $\{ G, {\cal H}_i\}$ are equivalent to 
the standard set, but form a true Lie algebra.  This result may be helpful in canonical 
quantization, but it has not yet been widely applied.

\subsection{Covariant formalism: first order form \label{seccc}}

As an alternative to the metric formalism of section \ref{secca}, we can write the
gravitational action in terms of an orthonormal frame (or ``tetrad,'' or ``vierbein'')
and a spin connection, treated as independent variables.  We introduce a coframe
$e_{\mu}{}^I$, where orthonormality means
\beq
g^{\mu\nu}e_{\mu}{}^Ie_{\nu}{}^J = \eta^{IJ} , \qquad
\eta_{IJ}e_{\mu}{}^Ie_{\nu}{}^J = g_{\mu\nu} .
\label{c25}
\eeq
Since parallel transport now requires a comparison of frames as well as tangent
spaces, we must introduce a new ``spin connection'' $\omega_\mu{}^I{}_J$,
with curvature $R_{\mu\nu}{}^{IJ}$, in addition to the usual Christoffel 
connection.  The gravitational action becomes
\beq
I = {1\over16\pi G} \int d^4x\, |\mathrm{det}\, e| 
   \left( e^{\mu I}e^{\nu J}R_{\mu\nu IJ}
  - 2\Lambda\right) + I_{\hbox{\scriptsize matter}} .
\label{c28}
\eeq
Along with diffeomorphism invariance, this action is invariant under local 
Lorentz transformations in the tangent space.  It is straightforward to show  
that extremizing this action reproduces the standard Einstein field 
equations \cite{Ashtekar}.   

A slightly different first-order formalism is useful as a starting point for loop 
variable quantization.  The dual of an antisymmetric two-index object $F_{IJ}$ 
is defined as
\beq
F^*_{IJ} = -{i\over2}\epsilon_{IJ}{}^{KL}F_{KL}
\label{c33}
\eeq
where the factor of $-i/2$ comes from the Lorentzian signature of spacetime and 
the requirement that $F^{**}=F$.  The spin connection determines a self-dual 
connection 
\beq
A_\mu{}^{IJ} = {1\over2}\left( \omega_\mu{}^{IJ} 
  - {i\over2}\epsilon^{IJ}{}_{KL}\omega_\mu{}^{KL}\right) \nonumber\\
\label{c34}
\eeq
and a corresponding curvature $F_{\mu\nu}{}^{IJ}$, where for the moment we 
work in {\em complex\/} general relativity.  Even though it involves only the 
self-dual part of the connection, the action
\beq
I = {1\over8\pi G} \int d^4x\, |\mathrm{det}\, e| 
  \left( e^{\mu I}e^{\nu J}F_{\mu\nu IJ}
  - \Lambda\right) + I_{\hbox{\scriptsize matter}} 
\label{c35}
\eeq
can be shown to yield the usual Einstein field equations 
\cite{Ashtekar,Jacobson,Samuel}.  

\subsection{Canonical formalism: first order form \label{seccd}}

The canonical form of the first-order action (\ref{c28}) gives little that is new
\cite{Peldan}.  The ADM description  of the self-dual 
action (\ref{c35}), on the other hand, is more interesting \cite{Ashtekar,Sen,%
Ashtekar2,Rovelli3,Gambini}.  Without loss of generality, we can first gauge-% 
fix the triad to require that $e^0{}_{{\hat I}}=0$ for ${\hat I} = 1,2,3$.  We 
now define a ``densitized triad'' and a local $\hbox{SO}(3)$ connection\footnote{%
Many authors use an $\hbox{SU}(2)$ spinorial description of the fields: 
${\tilde E}^i = 2{\tilde E}^i{}_{\hat I}\tau^{\hat I}$ and $A_i = 
A_{i{\hat I}}\tau^{\hat I}$, where $\tau^{\hat I} = -{i\over2}\sigma^{\hat I}$ 
are the spin $1/2$ $\mathit{su}(2)$ generators.  Note that normalizations vary.}
\beq
{\tilde E}^i{}_{\hat I} = \sqrt{q}\,e^i{}_{\hat I} , \qquad
A_i{}^{\hat I} = \epsilon^{0{\hat I}{\hat J}{\hat K}}A_{i{\hat J}{\hat K}} 
\label{c37}
\eeq
on the time slice $\Sigma_t$, and an $\hbox{SO}(3)$ curvature $F_{ij}{}^{\hat K}$
of $A_i{}^{\hat I}$. 

Using self-duality to write $A_\mu{}^{0{\hat I}}$ in terms of 
$A_\mu{}^{\hat I}$, we find that the action (\ref{c35}) becomes
\beq
I = {1\over8\pi G}\int dt\int_\Sigma d^3x\, \left[ 
   iA_i{}^{\hat I}\partial_t {\tilde E}^i{}_{\hat I} 
  - iA_{0{\hat I}}G^{\hat I} + iN^i{\cal V}_i
  - {1\over2}\left(N/\sqrt{q}\right){\cal S} \right],
\label{c40}
\eeq
with constraints 
\begin{eqnarray}
&&G^{\hat I} = D_j{\tilde E}^{j{\hat I}}  \nonumber\\
&&{\cal V}_i = {\tilde E}^j{}_{\hat I}F_{ij}{}^{\hat I}   \\
&&{\cal S} = \epsilon^{{\hat I}{\hat J}{\hat K}} 
       {\tilde E}^i{}_{\hat I}{\tilde E}^j{}_{\hat J}F_{ij{\hat K}} 
       - {\Lambda\over6}\eta_{ijk}\epsilon^{{\hat I}{\hat J}{\hat K}}
        {\tilde E}^i{}_{\hat I}{\tilde E}^j{}_{\hat J}{\tilde E}^k{}_{\hat K},
        \nonumber
\label{c41}
\end{eqnarray}
where $D_i$ is the gauge-covariant derivative with respect to the connection 
$A_i{}^{\hat I}$.  This action is already in canonical form, and we can read off 
the Poisson brackets,
\beq
\{ {\tilde E}^i{}_{\hat I}, A_j{}^{\hat J} \} 
  = - 8\pi i G\delta_{\hat I}^{\hat J}\delta^i_j{\tilde\delta}^3(x-x') .
\label{c42}
\eeq
The phase space is now that of an ordinary complex $\hbox{SO}(3)$
Yang-Mills theory---$A_i{}^{\hat I}$ is an $\hbox{SO}(3)$ ``gauge
potential'' and ${\tilde E}^i{}_{\hat I}$ is its conjugate ``electric
field''---with added constraints.  We can thus view general relativity 
as embedded in Yang-Mills theory.  In particular, the constraint $G^{\hat I} 
= 0$ is simply the Gauss law constraint, and generates ordinary $\hbox{SO}(3)$ 
gauge transformations, while ${\cal V}_i$ and $\cal S$  are analogs of the 
ADM momentum and Hamiltonian constraints, with an algebra that is 
essentially the surface deformation algebra (\ref{c23}).

In contrast to the ADM form, the constraints are now polynomial in both the 
positions and the momenta.  This simplification comes at a price, though: to 
define the self-dual connection we had to complexify the metric, and we must 
now  require that the metric and ordinary spin connection be real 
\cite{Ashtekar,Ashtekar3}.
 
The implementation of such ``reality conditions'' in the quantum theory has 
proven very difficult, and a good deal of work has been expended in trying to avoid 
them.  Note first that the need for a complex connection originated in the  factor 
of $i$ in the duality condition (\ref{c33}).  If our metric had Riemannian (positive 
definite) rather than Lorentzian signature, this $i$ would disappear.  This does 
not help much in the classical theory---spacetime is observably Lorentzian!---but 
as in ordinary quantum field theory, it may be that a ``Euclidean'' quantum theory 
can be ``Wick rotated'' back to Lorentzian signature \cite{Thiemann,Ashtekary}.

To explore another alternative, let us reexpress the spin connection in terms of 
ADM-like variables.  In our gauge $e^0{}_{{\hat I}}=0$, it is easy to check that
\beq
\omega_i{}^{0{\hat I}} = e^{j{\hat I}}K_{ij} = K_i{}^{\hat I} , \qquad
\Gamma_i{}^{\hat I} = {1\over2}\epsilon^{0{\hat I}{\hat J}{\hat K}}
  \omega_{i{\hat J}{\hat K}} ,
\label{c44}
\eeq
where $K_{ij}$ is the extrinsic curvature of $\Sigma_t$ and $\Gamma_i{}^{\hat I}$ 
is a three-dimensional $\hbox{SO}(3)$ connection, treated as a function of the 
triad.  The Ashtekar-Sen connection is then $A_i{}^{\hat I} =  \Gamma_i{}^{\hat I} 
+ i K_i{}^{\hat I}$.  Barbero and Immirzi have proposed a more general linear 
combination\cite{Barbero,Immirzi,Holst,Alexandrov},
\beq
A_i^{(\gamma)\hat I} =  \Gamma_i{}^{\hat I} + \gamma K_i{}^{\hat I} ,
\label{c47}
\eeq
where $\gamma$ is an arbitrary ``Immirzi parameter.''  The new 
connection has Poisson brackets
\beq
\{ {\tilde E}^i{}_{\hat I}, A_j^{(\gamma)\hat J} \} 
  = -8\pi \gamma G\delta_{\hat I}^{\hat J}\delta^i_j{\tilde\delta}^3(x-x') .
\label{c48}
\eeq
The resulting Hamiltonian constraint becomes considerably more
complicated,
\beq
{\cal S}^{(\gamma)} = \epsilon^{{\hat I}{\hat J}{\hat K}} 
       {\tilde E}^i{}_{\hat I}{\tilde E}^j{}_{\hat J}F_{ij{\hat K}} 
       -2{1+\gamma^2\over\gamma^2}
       {\tilde E}_{[{\hat I}}{}^i{\tilde E}_{{\hat J}]}{}^j
       (A_i^{(\gamma)\hat I} - \Gamma_i{}^{\hat I})
       (A_j^{(\gamma)\hat J} - \Gamma_j{}^{\hat J})
       - {\Lambda\over6}\eta_{ijk}\epsilon^{{\hat I}{\hat J}{\hat K}}
        {\tilde E}^i{}_{\hat I}{\tilde E}^j{}_{\hat J}{\tilde E}^k{}_{\hat K},
\label{c49}
\eeq
but as we shall see in section \ref{secfb}, it might still be manageable.  

\subsection{Gravity as a constrained $BF$ theory \label{secce}}

One more formulation of classical general relativity makes an appearance in 
quantization, particularly in the spin foam approach of section \ref{secfb}.  
Note first that the action (\ref{c28}) can be conveniently written in terms
of differential forms as
\beq
I = -{1\over64\pi G} \int \epsilon_{IJKL}e^I\wedge e^J\wedge \left(
  R^{KL} - {4\over3}\Lambda e^K\wedge e^L \right)  
  + I_{\hbox{\scriptsize matter}} ,
\label{c28a}
\eeq
where $e^I = e_\mu{}^I dx^\mu$ and $R^{KL} = R_{\mu\nu}{}^{KL}
dx^\mu\wedge dx^\nu$ are the tetrad one-form and curvature two-form.
In this notation, it is evident that the tetrad appears only in the antisymmetric 
combination $B^{IJ} = e^I\wedge e^J$.  We can trivially rewrite the action 
in terms of $B$, provided that we include a constraint that $B^{IJ}$ is of the
special form $e^I\wedge e^J$ for some $e^I$.  This constraint can be interpreted 
geometrically as a requirement that ``left-handed area,'' formed with the 
self-dual part of $B^{IJ}$, equal ``right-handed area,'' formed with the 
anti-self-dual part \cite{Reisenberger}.

The constraint can be implemented in a number of ways \cite{Capovilla}.  
For instance, (\ref{c28a}) is (almost) equivalent to the action
\cite{Reisenberger,Capovilla,Plebanski,DePietri0}
\beq
I = -{1\over64\pi G} \int \left[ \epsilon_{IJKL}
  B^{IJ}\wedge \left(R^{KL} - {4\over3}\Lambda B^{KL} \right) 
  + \phi_{IJKL}\left(B^{IJ}\wedge B^{KL} - e\epsilon^{IJKL}\right)\right] .
\label{c51}
\eeq
The value of this formalism comes in part from that fact that the unconstrained
action---the action (\ref{c51}) without the Lagrange multiplier $\phi_{IJKL}$---is 
that of a ``$BF$ theory,'' a topological theory with a well-understood quantization 
\cite{Birmingham}.   

\section{Covariant quantization}\setcounter{footnote}{0}

We turn at last to the problem of quantizing gravity.  I will begin with
``covariant quantization,'' quantization based on the four-dimensional 
action with no arbitrary choice of time.  With a single exception---covariant 
canonical quantization, to be discussed in section \ref{secec}---work on this 
program relies on perturbation theory of one kind or another.  Typically, the 
metric is split into a ``background'' ${\bar g}_{\mu\nu}$ and a ``quantum 
fluctuation'' $h_{\mu\nu}$,
\beq
g_{\mu\nu} = {\bar g}_{\mu\nu} +  \sqrt{16\pi G}\, h_{\mu\nu}
\label{d1}
\eeq
and quantities of interest are computed perturbatively in $h_{\mu\nu}$.  The 
background metric, in turn, can be determined self-consistently, for instance as 
an extremum of the quantum effective action.  The method earns the title 
``covariant'' because it yields quantities that are covariant under diffeomorphisms 
of the background metric.  Much of this field was pioneered by DeWitt 
\cite{DeWitt3,DeWitt4,DeWitt5,DeWitt6}, who introduced the background 
field method, developed Feynman rules,  and discovered (along with Feynman 
\cite{Feynman2}) the need for ghosts.

Although there are a number of ways to formulate the resulting theory, most  
begin with a path integral over either Lorentzian metrics or ``Wick rotated''  
Riemannian metrics.  I will describe two efforts to understand this path integral 
below.

\subsection{Lorentzian perturbation theory \label{secda}}

The most straightforward approach to quantizing gravity treats general relativity,
as much as possible, as an ordinary field theory, and is perhaps best understood as 
quantization of the massless spin two field of section \ref{secca}.  The starting 
point is the formal path integral
\beq
{}_{\hbox{\scriptsize out}}\!\left\langle 0 | 0 \right\rangle_{\hbox{\scriptsize in}}^J 
  = Z[J] = \int [dg] \exp\left\{ i(I_{\hbox{\scriptsize grav}} 
  + \int d^4x\, g_{\mu\nu}J^{\mu\nu}) \right\} ,
\label{d2}
\eeq
where $J^{\mu\nu}$ is an external source, included so that functional derivatives
of $Z[J]$ yield correlation functions:
\beq
\left({1\over i}{\delta\qquad \over\delta J^{\mu_1\nu_1}(x_1)}\dots
{1\over i}{\delta\qquad \over\delta J^{\mu_n\nu_n}(x_n)}Z[J]\right)\Biggl|_{J=0} 
  = {}_{\hbox{\scriptsize out}}\!\left\langle 0 |T g_{\mu_1\nu_1}(x_1)\dots
  g_{\mu_n\nu_n}(x_n)|0\right\rangle_{\hbox{\scriptsize in}} .
\label{d3}
\eeq
The {\em connected\/} Greens functions can be obtained as functional derivatives 
of $W[J] = -i\ln Z[J]$ (see, for example, chapter 2 of \cite{Buchbinder}).   The 
object is now to find $Z[J]$.

The generating functional (\ref{d2}) is a formal expression,  ordinarily  defined only 
in perturbation theory.  There are potential problems with such an approach---the 
time ordering in (\ref{d3}), for example, will be with respect to the background 
metric and may not express the ``true'' causality---but it is the best we can usually 
do.  The derivation of Feynman rules and a perturbative expansion is by now standard, 
although it is worth remembering that many of the subtleties were first discovered 
in early attempts to quantize gravity.

To obtain an expansion, we start with the decomposition (\ref{d1})
and change the functional integration variable from $g$ to $h$.  If the background 
metric satisfies the classical field equations, the terms linear in $h$ will drop out,
and the quadratic terms will be of the form 
\beq
I_{\hbox{\scriptsize grav}} = \int d^4x\sqrt{-{\bar g}}\,
 h_{\mu\nu}(D^{-1}[{\bar g}])^{\mu\nu\rho\sigma}h_{\rho\sigma} + \dots
\label{d4}
\eeq
In a simpler field theory, $D$ would be the propagator, and we could write down 
Feynman rules for interactions directly from the higher order terms in the
expansion.  For gravity, though, $D^{-1}$ has eigenfunctions with eigenvalue
zero, and is not invertible.  This is a direct consequence of diffeomorphism 
invariance:  if $h_{\mu\nu}$ is an infinitesimal diffeomorphism (\ref{c5}), 
the action is invariant, so such an $h_{\mu\nu}$ must be a zero mode of 
$D^{-1}$.  

As in gauge theories, the solution is to add a gauge-fixing term that breaks the 
invariance, and to restrict the path integral to gauge-fixed fields.   In quantum 
electrodynamics, one can simply insert the gauge condition into the path integral.  
But as Feynman first noticed \cite{Feynman2}, this leads to a loss of unitarity in 
quantum gravity, which must be compensated by adding extra ``ghost'' fields 
\cite{DeWitt5,Faddeev}.  We now understand that these ghosts arise from the 
geometry of the space of fields.  In fixing a gauge, we are changing variables from 
$h_{\mu\nu}$ to $\{{\tilde h}_{\mu\nu},\xi^\rho\}$, where ${\tilde h}_{\mu\nu}$ 
is a gauge-fixed field and $\xi^\rho$ parametrizes diffeomorphisms.  This change 
of variables involves a nontrivial Jacobian in the path integral, which can be 
expressed as a determinant of a differential operator $\cal D$; that determinant, 
in turn, can be written as a functional integral
\beq
\mathrm{det}{\cal D} = \int [d{\bar c}][dc]\exp\left\{
 i\int d^4x\,\sqrt{-\bar g}\, {\bar c}{\cal D} c \right\} ,
\label{d6}
\eeq
where $c$ and $\bar c$ are anticommuting bosonic fields.  This description is, 
of course, sketchy; for an elegant derivation based on the geometry of the 
space of fields, see \cite{Bern2}.

Once the gauge-fixing and ghost terms have been added, the path integral
determines Feynman rules with which one can compute correlation functions 
and other quantities of interest.  There are still subtleties, some not yet
resolved---the integration measure is not certain \cite{DeWitt2}, a globally 
valid gauge condition may not exist \cite{Gribov}, and off-shell quantities, 
while invariant under diffeomorphisms of ${\bar g}_{\mu\nu}$, can depend 
parametrically on the choice of gauge-fixing \cite{Vilkovisky}---but one can 
still begin to calculate.  In practice, one usually computes the effective action 
$\Gamma[{\bar g}]$, which is a functional Legendre transform of the generating 
function $W[J]$ and itself generates one-particle irreducible diagrams \cite{Abbott}.

The approach I have described treats gravity as an ordinary quantum field 
theory.  As in most such theories, computations often give divergent answers, 
and a key question is whether the infinities can be absorbed into redefinitions 
of the coupling constants---that is, whether the theory is renormalizable.  
Here the background field method is especially powerful, since any counterterms 
in the effective action must be diffeomorphism-invariant functions of 
${\bar g}_{\mu\nu}$; symmetry thus strictly limits the counterterms one 
must consider.

In 1974, 't Hooft and Veltman showed that pure quantum gravity was one-loop 
finite on shell, i.e., that at lowest order all counterterms were proportional to 
equations of motion and could be eliminated by field redefinitions \cite{tHooft}.  
They also showed, however, that the addition of a scalar matter field made the 
theory nonrenormalizable.  The next step for pure quantum gravity, the two-loop 
computation, was not carried out until 1986, when Goroff and Sagnotti finally 
showed that the theory was nonrenormalizable \cite{Goroff}:
a divergent counterterm  
\beq
\Gamma^{(2)}_{\hbox{\scriptsize div}} = {209\over2880(4\pi)^2}
 {1\over\epsilon} \int d^4x\sqrt{-g}\, R^{\mu\nu}{}_{\pi\rho}
 R^{\pi\rho}{}_{\sigma\tau}R^{\sigma\tau}{}_{\mu\nu} 
\label{d7}
\eeq
appears in the effective action.  The result was confirmed by van de
Ven \cite{Ven}, and effectively put an end to the hopes of a conventional
quantum field theoretical approach to gravity.

There are several ways one may react to this failure of renormalizability:
\begin{enumerate}
\item Perhaps the right combination of matter fields can cancel the divergences.  
Fermion and boson loops contribute with opposite signs, and supersymmetry 
forces some exact cancellations; indeed, symmetry considerations rule out terms 
like (\ref{d7}) in supergravity.  Unfortunately, using new computation techniques 
inspired by string theory, Bern et al.\ have now shown that supergravity is also 
nonrenormalizable \cite{Bern}, although the case of maximal ($N=8$) 
supersymmetry is not completely settled.
\item Perhaps the action (\ref{c1}) should be revised, for instance by including 
terms involving higher powers of the curvature.  Even if they are not present 
in the original action, such terms will appear in the effective action when one 
renormalizes the stress-energy tensor \cite{Utiyama,Birrell}.  A typical higher
order action would take the form
\beq
I_{\hbox{\scriptsize grav}}^{\hbox{\scriptsize (quad)}} = \int d^4x\sqrt{-g}
  \left[ {1\over16\pi G}(R - 2\Lambda) + aR^2 
  + b C_{\mu\nu\rho\sigma}C^{\mu\nu\rho\sigma}\right] 
\label{d8}
\eeq
where $C_{\mu\nu\rho\sigma}$ is the Weyl tensor.  The quadratic curvature 
terms generically suppress divergences, since they lead to a propagator that goes  
as $1/k^4$ rather than $1/k^2$ at large momenta.  Indeed, Stelle has shown that 
the action (\ref{d8}) is renormalizable for most choices of the coupling constants 
\cite{Stelle,Buchbinder}.  Unfortunately, though, the resulting quantum theory is 
also perturbatively nonunitary: a typical propagator has the form
\beq
{m^2\over k^2(k^2+m^2)}  = {1\over k^2} - {1\over k^2 + m^2} ,
\label{d9}
\eeq
and the minus sign in the second term indicates a negative norm ``ghost'' state.  
The coupling constants in (\ref{d8}) can be adjusted to eliminate ghosts, but at 
precisely those values, the theory again becomes nonrenormalizable.  It may be 
that a full nonperturbative treatment restores unitarity \cite{Tomboulis,Kay}, 
but this idea remains speculative, and the restoration appears to come at the 
expense of causality at short distances \cite{Lee}.  
\item Perhaps when the perturbation series is properly summed, quantum
gravity is {\em finite}.  The problem with a nonrenormalizable theory is that
the effective action contains infinitely many terms, each with an undetermined
coupling constant, thus drastically limiting predictive power.  But if the higher
order terms have finite coefficients that can be determined from the original 
Einstein-Hilbert action, the problem largely disappears.  The idea that quantum 
gravity eliminates divergences gains support from the observation that classical 
gravity eliminates the infinite electromagnetic self-energy of a charged point 
particle \cite{ADM2}, and from various partial resummations of Feynman 
diagrams \cite{DeWitt,Khriplovich,Isham2} and similar approximations 
\cite{Ohanian,Barvinsky,Barvinsky2}.  But we do not know how to perform a 
complete summation of perturbation theory, of course, so this argument really
points toward the need for new nonperturbative methods.   
\item Perhaps we have used the wrong set of variables.  The metric variables 
in (\ref{d2}) are not diffeomorphism-invariant, and the correlation functions 
(\ref{d3}) are actually correlators of complicated nonlocal functions of curvatures 
determined by the gauge-fixing \cite{Tsamis}.  It might be more sensible to start 
with invariant fields.  Now, $S$-matrix elements are unchanged by local field 
redefinitions, but nonlocal redefinitions can change the theory. Attempts to do 
perturbation theory using invariant fields have so far only made the divergences 
worse, but we may not yet know the right choices.   
\item Perhaps we are doing perturbation theory wrong.  The split (\ref{d1})
of the metric into a background piece and a fluctuation cannot be quite right, 
since for $h_{\mu\nu}$ large enough the metric will no longer have Lorentzian
signature.  The path integral is missing the geometry of the space of metrics;
this may be a fatal problem \cite{DeWittx}.  Or perhaps we need to expand
in a different parameter: the three-dimensional Gross-Neveu model, for
example, is nonrenormalizable in the weak coupling expansion but renormalizable 
in the $1/N$ expansion \cite{Gross2}.
\item Perhaps we can live with nonrenormalizability.  Recall that in quantum
field theory, the ``coupling constants'' in the effective action really depend on
energy scale, with a flow given by the renormalization group \cite{Weinberg}. 
Flows that avoid unphysical high energy singularities form ``ultraviolet critical 
surfaces'' in the space of coupling constants.  Weinberg calls a theory asymptotically 
safe if its coupling constants lie on such a surface, and proposes that quantum 
gravity might be such a theory \cite{Weinberg2}.  If the relevant surface turns 
out to be finite dimensional, we gain the same advantage we would have with a 
finite theory: although the number of coupling constants is infinite, all are 
determined in terms of a finite number of independent parameters.
\item Perhaps the perturbative approach is simply wrong, either because the metric 
is not a fundamental field (the approach of string theory) or because the expansion 
around a smooth classical background is invalid (the approach of loop quantum 
gravity).
\end{enumerate}

Note that even though the perturbation theory described here does not provide an 
ultimate quantum theory of gravity, it can still provide a good {\em effective\/} 
theory for the low energy behavior of quantum gravity \cite{Weinberg2,Donoghue}.  
Whatever the final theory, gravity at low energies is at least approximately described 
by a massless spin two field, whose action must look like the Einstein-Hilbert action 
plus possible higher order terms.  If we restrict our attention to processes in which 
all external particles have energies of order $E\ll M_{\hbox{\scriptsize Planck}}$, 
we can write an ``effective action'' that includes all local terms allowed by 
diffeomorphism invariance.\footnote{This use of the term ``effective action'' 
differs from the previous definition as a Legendre transform of the generating 
functional.  Both usages are common.}  Physically, the uncertainty principle 
guarantees that any high energy intermediate states involve short distances, and 
can thus be described by a local action, much as Fermi theory approximates 
electroweak theory at energies far enough below the mass of the $W$ boson.  

If we now use this effective action to compute low energy, long distance processes, 
we find that high energy corrections from higher order terms will be suppressed by 
powers of  $E/M_{\hbox{\scriptsize Planck}}$.  Low energy quantum effects can be 
isolated, and give, for example, modifications to the long distance Newtonian limit
\cite{Donoghue2}:
\beq
V(r) = -{Gm_1m_2\over r}\left[ 1 - {G(m_1 + m_2)\over rc^2} 
  - {127\over30\pi^2}{G\hbar\over r^2c^3} + \dots \right] .
\label{d9a}
\eeq

\subsection{The Euclidean path integral \label{secdb}}

The path integral (\ref{d2}) was designed to compute correlation functions, but 
it has another use as well.  Let $M$ be a manifold with boundary, and fix the spatial 
metric $q_{ij}$ on $\partial M$.  Then the path integral over metrics on $M$ with 
the specified boundary data should give a transition amplitude.  For example, if 
$\partial M$ has two disconnected components $\Sigma$ and $\Sigma'$, the path 
integral determines a transition amplitude between three-geometries $q_{ij}$ 
and $q_{ij}'$; if $\partial M$ has three disconnected components, the path integral 
gives an amplitude for the creation of a ``baby universe''; and if $M$ has only a single 
boundary, the path integral describes ``tunneling from nothing'' and provides a 
candidate for the wave function of the Universe, the Hartle-Hawking wave function 
\cite{Hartle}.  The problem of nonrenormalizability remains, of course, but even if 
the Einstein-Hilbert action only describes an effective field theory, we may still be 
able to obtain useful approximate information.

\begin{figure}
\begin{picture}(100,140)(-150,10)
\qbezier(21,20)(0,56)(16,100)
\qbezier(95,20)(116,56)(102,102)
\qbezier(21,20)(58,-20)(95,20)
\qbezier(16,100)(20,120)(12,140)
\qbezier(102,102)(98,121)(106,140)
\qbezier(12,140)(60,115)(106,140)
\qbezier(12,140)(60,165)(106,140)
\put(54,136){$\Sigma$}
\put(54,50){$M$}
\end{picture}
\caption{A manifold $M$ with a single boundary $\Sigma$ describes the
birth of a universe in the Hartle-Hawking approach to quantum cosmology.
\label{fig2}}
\end{figure}
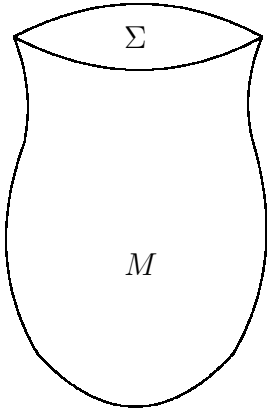
In this approach, we can discard the external source $J$ in (\ref{d2}), and 
consider the path integral as a function of the boundary data $q_{ij}$.  The 
leading contribution will come from ``saddle points,'' classical solutions of 
the field equations with the prescribed boundary data.  For most topologies, 
though, no such classical solutions exist.  The Hartle-Hawking path integral, 
for instance, requires a Lorentzian metric on a four-manifold $M$ with 
a single spacelike boundary; but simple topological arguments show that most 
manifolds $M$ admit no such metric \cite{Sorkin2,Reinhart}.  More generally, 
a four-manifold $M$ admits a Lorentzian metric that mediates topology change 
between closed spacelike boundaries only if its Euler number is zero, and even 
then there can be dynamical obstructions to topology change \cite{Tipler,Borde}.  

These obstructions can be eased in several ways---for instance, by 
allowing mild singularities \cite{Horowitz}---but the usual choice is to
``analytically continue'' to Riemannian signature \cite{Hawking2};
hence the term ``Euclidean path integral.''  This choice is inspired by
the analogy to ordinary quantum field theory, where Euclidean saddle points, 
or instantons, give a good semiclassical description of tunneling \cite{Coleman}.  
Of course, the relationship between Wick-rotated correlation functions in 
quantum field theory and their Lorentzian counterparts is rigorously understood, 
while no such result is known for quantum gravity; the Euclidean path integral 
should really be thought of as a {\em definition\/} of a quantum theory.

One problem with this definition is that the Euclidean action $I_E[g]$---that is, 
the action (\ref{c1}) evaluated on metrics with Riemannian signature---is not 
positive definite.  The problem lies in the ``conformal factor'': if we consider 
a new metric ${\tilde g}_{\mu\nu} = \Omega^2g_{\mu\nu}$, we find that
\beq
I_E[{\tilde g}] = -{1\over16\pi G}\int d^4x\sqrt{g}\left[ \Omega^2 R[g] 
  - 2\Omega^4\Lambda + 6 g^{\mu\nu}\partial_\mu\Omega\partial_\nu\Omega
  \right] ,
\label{d10}
\eeq
so by allowing $\Omega$ to vary fast enough we can make the action arbitrarily
negative.  This is no problem at the classical level, of course, and the unwanted 
conformal term is exactly canceled by the Faddeev-Popov determinant at one 
loop \cite{Mazur2} and perhaps more generally \cite{Dasgupta}.  For the full 
path integral, though, the situation is not entirely clear, and might depend crucially
on the integration measure.  Metrics with large negative actions may ultimately be
unimportant---by analogy, we can evaluate $\int_0^1 dx/\sqrt{x}$ even though 
the integrand blows up at $0$---but it may alternatively be necessary to further 
``analytically continue'' the conformal factor \cite{Hawking2}.  Once we permit 
such a step, though, the choice of an integration contour in the space of complex 
metrics becomes ambiguous, and different choices may lead to different 
amplitudes \cite{Louko}.

The Euclidean path integral depends on the three-metric $q_{ij}$ and the
corresponding boundary data for matter, but also on the topology of the interpolating
four-manifold $M$.  We must thus decide which four-manifolds to include.
In the absence of any natural way to single out a preferred topology, and
following Wheeler's arguments for fluctuating topology \cite{Wheeler}, 
the usual choice is to sum over all manifolds, thus allowing for ``spacetime 
foam.''  For simplicity, let us specialize to the case of a single boundary; the 
Hartle-Hawking wave function is then
\beq
\Psi[q,\varphi|_\Sigma] = \sum_M \int [dg][d\varphi] e^{-I_E[g,\varphi;M]}
  \approx \sum_M \Delta_M[{\bar g},{\bar\varphi}] 
  e^{-I_E[{\bar g},{\bar\varphi};M]}
\label{d11}
\eeq
where $I_E$ is the Euclidean action, $\varphi$ is a generic collection of matter 
fields, $\bar g$ and $\bar\varphi$ are the classical solutions with the specified 
boundary data $q$ and $\varphi|_\Sigma$, and $\Delta_M$ is the one-loop 
correction, which includes the Faddeev-Popov determinant (\ref{d6}) and 
determinants coming from integrating quadratic fluctuations around 
$\{\bar g,\bar\varphi\}$ \cite{Hawking3,Christensen}.   

Gibbons and Hartle have argued that since the Universe is now Lorentzian, 
the relevant saddle points of (\ref{d11}) are ``real tunneling geometries,'' 
which extrapolate from Riemannian metrics in the early Universe to Lorentzian
metrics later \cite{Gibbons2}.  If one views the boundary $\Sigma$ in figure
\ref{fig2} as the site of such a signature change, smoothness of the metric 
requires that $\Sigma$ have vanishing extrinsic curvature.  From the point of 
view of quantum theory, this restriction doesn't quite make sense, though: to
define a wave function on $\Sigma$ we must allow the induced metric $q_{ij}$
to vary, and we cannot simultaneously specify the extrinsic curvature, which
is conjugate to the metric.  It may be shown, however, that the wave function
$\Psi[q]$, viewed as a functional of $q$, is {\em extremal\/} when the spatial 
metric is the boundary value of a real tunneling metric.  Moreover, if we fix
``time'' by specifying $\mathrm{Tr}\,K = 0$, it is likely that these extrema 
are in fact maxima, so real tunneling geometries emerge as (probabilistic) 
predictions of the Euclidean path integral rather than extra inputs \cite{Carlip3}.

The Euclidean path integral gives us a wave function that can at least formally
be shown \cite{Hartle2} to satisfy the Wheeler-DeWitt equation of canonical 
quantum gravity (see section \ref{secea}).  The interpretation of this wave 
function is not entirely clear---in particular, it contains no explicit reference 
to time---and will be discussed in section \ref{sece}.  One advantage of the 
path integral formalism is that it suggests a natural inner product on the space 
of such wave functions: if $\Psi$ and $\Phi$ are obtained from path integrals 
over manifolds $M_1$ and $M_2$ with diffeomorphic boundaries $\Sigma$, 
we can ``glue'' the two manifolds at the boundary to obtain a closed manifold 
$M=M_1\cup_\Sigma M_2$, and define
\beq
\langle\Phi|\Psi\rangle = \int [dg][d\varphi]e^{-I_E[g,\varphi;M]} .
\label{d12}
\eeq

With the exception of some recent work string theory and ``spin foams,'' the 
Euclidean path integral is the only standard approach to quantum gravity in which 
fluctuations of spacetime topology appear naturally.  It is not entirely clear what  
the sum over topologies in (\ref{d11}) means, since four-manifolds are not 
classifiable \cite{Geroch}, and one might wish to sum over spaces more 
general than manifolds \cite{Hartle0,Schleich0}.  Nevertheless,  we can get 
some idea of the role of complicated topologies by considering the saddle point 
approximation.

In the absence of matter, a simple computation of the classical action gives
\beq
e^{-I_E[{\bar g};M]} = \exp\left\{{9\over8\pi \Lambda G}{\tilde v}(M)\right\} ,
\label{d13}
\eeq
where $\tilde v$ is the ``normalized volume'' of $M$, obtained by rescaling the
metric to set the scalar curvature to $\pm 12$.   For a negative cosmological
constant, ${\tilde v}(M)$ is a good measure of topological complexity; for 
hyperbolic manifolds, for example, it is proportional to the Euler number.  At 
first sight, (\ref{d13}) thus implies that the path integral is dominated by the 
simplest topologies, with contributions from more complex topologies 
exponentially suppressed.  On the other hand, there are a great many complex, 
high $\tilde v$ topologies; in fact, their number increases faster than exponentially, 
and arbitrarily complicated manifolds dominate both the partition function 
\cite{Carlip4} and the Hartle-Hawking wave function \cite{Ratcliffe}.  The full
sum over topologies is actually badly divergent, perhaps indicating that a universe 
with $\Lambda<0$ is ``thermodynamically'' unstable.

For $\Lambda>0$, the manifold with the largest normalized volume is the
four-sphere, which gives the largest single contribution to the sum (\ref{d11}).  
Much less is known about the overall  sum; the relevant mathematics does not 
yet exist.  As Baum \cite{Baum} and Hawking \cite{Hawking4} first noted, 
however, the exponent (\ref{d13}) is infinitely peaked at $\Lambda=0$, so if the 
cosmological constant is somehow allowed to vary, small values of $\Lambda$ will 
be strongly preferred.  These ideas were elaborated by Coleman \cite{Coleman2}, 
who considered ``approximate instantons'' consisting of large spherical regions 
connected by small wormholes; he argued that the sum over such configurations 
exponentiates the exponential in (\ref{d13}), giving a wave function  even more
sharply peaked at $\Lambda=0$.  As an explanation for the observed absence of 
a cosmological constant, this picture has the right ``flavor''---wormholes with 
very small mouths can connect very distant points, thus coupling the small scale 
quantum field theory that can produce a large vacuum energy with the large scale 
cosmology at which $\Lambda$ is observed to be nearly zero.  But the argument 
assumes a good deal about quantum gravity that is not  well understood---for 
example, the correct combinatorics of wormholes is not clear \cite{Carlip5}---and 
it is plagued by problems such as a failure to suppress (unobserved) ``giant wormholes'' 
\cite{Polchinski2}.

\subsection{Consistent histories \label{secdc}}

While the path integral is often presented as an alternative to canonical quantization, 
typical path integral approaches implicitly rely on canonical methods.  In the path
integral (\ref{d2}), for instance, the vacuum states $| 0 \rangle_{\hbox{\scriptsize in}}$ 
and $| 0\rangle_{\hbox{\scriptsize out}}$ are defined only through canonical
quantization, and quantities such as propagators only become physically meaningful 
when suitable initial and final states are folded in.  The Hartle-Hawking path integral 
(\ref{d11}) defines a state---possibly even one for each four-manifold $M$---and
perhaps an inner product, but it does not give us the full Hilbert space or the set of 
operators that act on states.

The consistent histories (``decoherent histories,'' ``post-Everett quantum mechanics'') 
program \cite{Hartlex} is in part an attempt to define a generalized quantum mechanics 
directly in terms of path integrals, without reference to states or operators.  The 
fundamental objects are histories---for general relativity, spacetimes---or, more 
generally,  coarse-grained histories, described only incompletely by a temporal 
sequence of alternatives.  We know from quantum mechanics that histories cannot 
always be assigned consistent probabilities, but the members of a set of sufficiently 
decoherent ``consistent histories'' can; such histories have negligible interference, and 
their probabilities obey the usual axioms of probability theory.

The consistent histories program is not an approach to quantum gravity per se, but 
rather a generalization of standard quantum mechanics that is potentially useful 
for quantum gravity.  Ordinary quantum mechanics arises as a special limit, but the 
new formalism is broader.  In particular, by dropping the usual requirement of a 
fixed foliation of spacetime, it may evade the ``problem of time'' and allow a 
quantum mechanical treatment of situations such as topology change in which no 
global time slicing exists. 

\section{Canonical quantization \label{sece}}\setcounter{footnote}{0}

To quantize gravity, one must choose between two competing geometric
structures: the diffeomorphism-invariant four-geometry of spacetime 
and the infinite-dimensional symplectic geometry of the space of metrics 
and momenta on a fixed time slice.  In the preceding section, we took the 
former as fundamental.  In this section, we focus on the latter.  

There is no universal prescription for canonically quantizing a classical theory.
Roughly speaking, we would like to promote Poisson brackets to commutators 
of operators on a Hilbert space:
\beq
\{x ,y \} \rightarrow {1\over i\hbar}[{\hat x},{\hat y}] .
\label{e0}
\eeq
Ambiguities arise because this is not, strictly speaking, possible: even for 
the simplest classical theories, factor ordering difficulties prevent us from 
simultaneously making the substitution (\ref{e0}) for all classical variables 
\cite{VanHove}.  Instead, we must pick a subalgebra of ``elementary classical 
variables''---usually, though not always, phase space coordinates 
$(p,q)$---and construct the remaining operators from elements of this set 
\cite{Isham3}.

The requirement that the resulting operator algebra be irreducible adds 
further restrictions.  In particular, if the classical variables $(p,q)$ have 
the usual Poisson brackets, we must choose a polarization, a splitting of 
phase space into ``positions'' and ``momenta'' so that wave functions 
depend on only half the variables.  There is sometimes a natural choice of 
polarization---if the phase space is a cotangent bundle, for example, we 
can choose the base space as our configuration space and the cotangent vectors 
as momenta---but often there is not, and there is no guarantee that different 
polarizations yield equivalent quantum theories.  

Further technical issues arise if the topology of phase space is nontrivial.  Even
more care is needed for the proper treatment of fields: field operators are 
distribution-valued, with  products that are ill-defined without regularization, 
and the conventional Hilbert space formalism must be generalized.   I will 
generally avoid explicit discussion of these issues, though they may ultimately 
prove crucial.

Our starting point for gravity is the canonical formalism of section \ref{seccb}.  
Its fundamental feature is the existence of the constraints (\ref{c16})--(\ref{c17}), 
and the first question we must ask is how to enforce them in a quantum theory.  
Broadly speaking, we have two choices: we can quantize first and then impose 
the constraints as conditions on wave functions (Dirac quantization), or we can 
first eliminate the constraints to obtain a classical ``physical phase space'' and 
then quantize (reduced phase space quantization).  The two methods have been 
applied to simple models, and have been found to often be inequivalent 
\cite{Ashtekar4,Romano,Schleich,Epp}.  There is, unfortunately, no compelling 
reason to prefer one or the other for quantum gravity.

\subsection{Dirac quantization and the Wheeler-DeWitt equation \label{secea}}

We begin with Dirac quantization (``quantize, then constrain'') \cite{Dirac2,%
DeWitt7}.  In this approach, quantization of gravity takes place in a series
of steps:
\begin{enumerate}
\item Define an auxiliary Hilbert space $H^{\hbox{\scriptsize (aux)}}$ 
consisting of a suitable collection of functions (more generally, sections of a
bundle) $\Psi[q]$ of the ``positions'' $q_{ij}$.
\item Promote the canonical Poisson brackets (\ref{c14}) to commutators,
\beq
[ q_{ij}(x), \pi^{kl}(x') ] = {i\over2}(\delta_i^k\delta_j^l
  + \delta_i^l\delta_j^k)\tilde\delta^3(x-x')  ,
\label{e1}
\eeq
and represent the canonical momenta as operators,
\beq
\pi^{kl} = -i{\delta\  \over\delta q_{kl}} .
\label{e2}
\eeq
\item Write the constraints as operators acting on $H^{\hbox{\scriptsize (aux)}}$,
and demand that physical states be annihilated by these operators.
\item Find a new inner product on the space of states annihilated by the
constraints, and form a new physical Hilbert space $H^{\hbox{\scriptsize (phys)}}$.
Note that operators on $H^{\hbox{\scriptsize (aux)}}$ will take physical states
to physical states only if they commute with the constraints.
\item Figure out how to interpret the resulting wave functions and operators.
\end{enumerate}
While these steps are easy to sketch, none is easy to accomplish:

{\bf Step 1}: We must decide exactly what functions to include in 
$H^{\hbox{\scriptsize (aux)}}$, and what spatial metrics $q_{ij}$ to allow as 
arguments.  Must $q_{ij}$ be nonsingular?  If our spatial hypersurface is open, do 
we restrict the asymptotic behavior of the metric?

{\bf Step 2}: The commutators (\ref{e1}) may be a bad starting point, since
they do not respect positivity of the spatial metric\cite{Isham3}.  Just as the
momentum $p$ generates translations of the position $q$ in ordinary quantum 
mechanics, $\pi^{ij}$ generates translations of $q_{ij}$, and nothing forbids
a translation to negative values.  One possible solution is to use triads rather than 
metrics as  ``positions,'' since there is no need for $e_i{}^I$ to be positive \cite{Isham4}.  
Another is to replace the commutators (\ref{e1}) with affine commutators of 
the form $[q, q\pi]\sim iq$, $[q\pi,q\pi]\sim iq\pi$, since these can be shown to 
preserve positivity \cite{Klauder}.

{\bf Step 3}: From (\ref{e2}), the momentum constraints (\ref{c17}) become 
operator equations
\beq
2i\, {}^{(3)}\nabla_j\left( {\delta\Psi[q] \over\delta q_{ij}}\right)  = 0 .
\label{e3}
\eeq
In principle, these are easy:  as in the classical case, the constraints 
generate spatial diffeomorphisms, and a state $\Psi[q]$ satisfies (\ref{e3}) 
precisely when it is a diffeomorphism-invariant functional of $q_{ij}$.  
Equivalently, we can consider $\Psi$ to be a function on ``superspace,'' the 
quotient $\mathrm{Riem}(\Sigma)/\mathrm{Diff}_0(\Sigma)$ of 
Riemannian metrics on $\Sigma$ modulo diffeomorphisms \cite{Wheeler,%
Wheeler3,Ebin,Fischer}.\footnote{$\mathrm{Diff}_0(\Sigma)$ is the 
identity component of the group of diffeomorphisms of $\Sigma$, that is, 
the group of diffeomorphisms that can be built up from infinitesimal
transformations.  The ``large'' diffeomorphisms---those that cannot
be continuously deformed to the identity---are not generated by the
momentum constraints, and may act as symmetries rather than gauge
transformations; see section \ref{seceb}.}

The Hamiltonian constraint (\ref{c16}) is harder.  It takes the operator form
\beq
\left[ (16\pi G)G_{ijkl}
  {\delta\ \over\delta q_{ij}}{\delta\ \over\delta q_{kl}}
  + {1\over16\pi G}\sqrt{q}\,({}^{(3)}\!R - 2\Lambda)\right]\Psi[q] = 0 ,
\label{e4}
\eeq
where
\beq
G_{ijkl} = {1\over2}q^{-1/2}\left( q_{ik}q_{jl} + q_{il}q_{jk} -
  q_{ij}q_{kl}\right)
\label{e5}
\eeq
is DeWitt's supermetric.  Equation (\ref{e4}) is the famous (or notorious) 
Wheeler-DeWitt equation \cite{DeWitt7,Wheeler3}.

We cannot, of course, find the general exact solution to (\ref{e4}).  Before
we even start, it is worthwhile to consider some of the difficulties:
\begin{itemize}
\item The equation contains a product of two functional derivatives at the
same point, whose action on a functional of $q_{ij}$ will typically give a
factor $\delta^3(0)$.  The Wheeler-DeWitt equation must be regularized.
\item The operator orderings in (\ref{e3}) and (\ref{e4}) are ambiguous.  With 
the orderings given, the commutator algebra of the constraints does not close
properly.  In general, closure of the algebra cannot be separated from the choice
of regularization \cite{Tsamis2,Friedman2}.
\item Unless one works with functionals that are {\em explicitly\/} invariant
under spatial diffeomorphisms, the coupling of the Wheeler-DeWitt equation and 
the momentum constraints causes severe problems \cite{Carlip6}, typically in the 
form of nonlocal terms.  The alternative of working with invariant expressions 
from the beginning is not much easier, since one must include nonlocal invariants 
such as $\int R\Delta^{-1}R$ \cite{Banks}.
\item It is not clear what, if any, boundary conditions we need.
\end{itemize}
Despite these problems, valiant efforts have been made to solve the
Wheeler-DeWitt equation.  One approach is to freeze out all but a few degrees of 
freedom to form a ``minisuperspace''  (see section \ref{secha}).  Another is to 
search for a perturbative expansion, in powers of the inverse cosmological constant 
\cite{Banks} or the inverse Planck mass \cite{Kiefer,Barvinsky4}.  One very 
nice result---see \cite{Kiefer} for a review---is that a Born-Oppenheimer 
approximation to the Wheeler-DeWitt equation for gravity coupled to matter 
yields the Schr{\"o}dinger equation for matter as a first approximation, with 
quantum gravitational corrections appearing at the next order.  A Feynman
diagram approach to higher order corrections exists, but it seems that new 
ultraviolet divergences appear and that the problem of nonrenormalizability 
reemerges \cite{Barvinsky4}.  

{\bf Step 4}:  Suppose that we have somehow found a set of solutions of the 
Wheeler-DeWitt equation.  We must now choose an inner product and make 
this into a Hilbert space.  The DeWitt supermetric (\ref{e5}) has signature 
$( -+++++ \phantom.\!)$, so the Wheeler-DeWitt equation formally resembles a 
Klein-Gordon equation on superspace.  A natural guess for a Klein-Gordon-like 
inner product is therefore, schematically \cite{DeWitt7},
\beq
\langle\Psi|\Phi\rangle = {1\over2i} \prod_{x\in\Sigma}
  \int_S d\Sigma^{ij} G_{ijkl}\,
  \Psi^*[q]{ {{\leftrightarrow}\atop{\displaystyle\delta}}\ 
  \over\delta q_{kl} }\Phi[q]
\label{e6}
\eeq
where $S$ is a hypersurface in superspace with directed surface element
$d\Sigma^{ij}$.  Like the inner product in Klein-Gordon theory, though, (\ref{e6}) 
is not positive definite, and cannot be interpreted directly as a probability.  In 
ordinary Klein-Gordon theory, this problem is solved by restricting to positive 
energy solutions.  That seems to make little sense in the context of quantum
gravity (though see \cite{Wald2}), where the analogous solutions correspond 
roughly to expanding universes; and even with such a restriction, the inner 
product (\ref{e6}) still generally fails to be positive definite \cite{Kuchar}.  
Moreover, the inner product (\ref{e6}) vanishes for real wave functions, while 
as Barbour has stressed \cite{Barbour}, the Wheeler-DeWitt equation is real, 
and hence does not couple real and imaginary parts of $\Psi[q]$ in any natural way.

An alternative \cite{Hartle} is to take the inner product to be simply
\beq
\langle\Psi|\Phi\rangle = \int [dq] \Psi^*[q]\Phi[q]  .
\label{e7}
\eeq
But here we must directly confront the ``problem of time.''  In classical
canonical gravity, the three-metric $q_{ij}$ on a hypersurface $\Sigma$ 
contains information not only about the gravitational degrees of freedom
(the usual two transverse polarizations for weak fields), but also about the 
``time'' on $\Sigma$, that is, the way $\Sigma$ fits into a four-manifold 
\cite{Baierlein}.  As Wheeler and his collaborators have stressed, one must 
pick out a time variable before a probability interpretation makes sense.  If
one integrates over the full three-metric in (\ref{e7}), the norm implicitly 
includes an integration over time, and must diverge \cite{Isham0,Kuchar,Wald3}.  

Woodard has argued that this problem arises because we have not ``gauge-fixed
the inner product'' \cite{Woodard}. The Hamiltonian constraint does not 
merely constrain fields; it also generates transformations that leave physical 
states invariant, and these must somehow be factored out when one forms
an inner product.  This argument gains support from the path integral inner 
product of section \ref{secdb}, and is known to be correct for certain simpler 
field theories \cite{Witten}.  A similar but more mathematically sophisticated 
approach, ``refined algebraic quantization,'' is currently the subject of considerable 
attention \cite{Marolf,Giulini,Landsman}.

Once we have an inner product, we must also find self-adjoint operators that 
carry solutions of the Wheeler-DeWitt equation to solutions.  The obvious 
candidates fail: $q_{ij}$, for example, is not invariant under spatial diffeomorphisms.  
Torre has shown that classical physical observables\footnote{By ``observables'' I 
mean functions that have vanishing Poisson brackets with the constraints, or 
operators that commute with the constraints.  This is a loaded term; Kucha{\v r},
for example, argues that observables ought not be required to commute with
the Hamiltonian constraint \cite{Kuchar3}.} cannot be local functionals of 
$(q_{ij},\pi^{ij})$ \cite{Torre}.  Kucha{\v r} has further shown that there are 
no classical observables, even nonlocal, that are linear in the momenta 
\cite{Kuchar4}.

{\bf Step 5}: Finally, if we have succeeded in the formal construction of a Hilbert 
space of solutions of the Wheeler-DeWitt equation, we must interpret our 
results.  Some of the problems of interpretation have been touched on above: our 
observables will be nonlocal, and it is not clear what we will mean by ``time'' or 
``evolution.''  Beyond these issues, we will have to decide what a ``wave function 
of the Universe'' means, a question that touches upon deep uncertainties in the 
interpretation of quantum mechanics \cite{Bell,Hartlex,Wheeler5,Page2,Smolin} 
and goes far beyond the scope of this review.

Some of these problems may be alleviated by a slightly different approach to 
Dirac quantization, the ``functional Schr{\"o}dinger representation''
\cite{Bergmannx,Teitelboimx,Kuchar5}.  In this approach, we isolate a choice of 
time before quantizing, by choosing a classical ``time function'' $T[q,\pi;x]$ built 
from the metric and momenta.  A canonical transformation makes $T$ and 
its conjugate $\pi_T$ into phase space coordinates: $(q,\pi) \rightarrow 
(T,\pi_T;{\tilde q},{\tilde\pi})$.  If we then solve the classical Hamiltonian constraint 
for $\pi_T$,
\beq
\pi_T = - h_T[{\tilde q},{\tilde\pi},T],
\label{e7a}
\eeq
the analog of the Wheeler-DeWitt equation becomes
\beq
i{\delta\ \over\delta T}\Psi[{\tilde q},T] 
  = h_T[{\tilde q},-i{\delta\ \over\delta{\tilde q}},T]\Psi[{\tilde q},T] .
\label{e7b}
\eeq

This constraint looks more like an ordinary Tomonaga-Schwinger ``many-fingered 
time'' Schr{\"o}dinger equation, and one can define a Schr{\"o}dinger inner product 
like (\ref{e7}) restricted to fixed $T$.  But the problems now reappear in a different 
guise \cite{Kuchar,Isham2b}:
\begin{itemize}
\item For many spatial topologies, including the simplest ones, there are global
obstructions to finding any solution (\ref{e7a}) of the Hamiltonian constraint 
\cite{Hajicek,Hajicek2,Torre2}: the topology of phase space may preclude the 
existence of a well-behaved global time function. Moreover, even locally, most 
simple choices of time function $T$ lead to physically unacceptable descriptions
in which the ``time'' at a given event depends not only on the event, but on the 
choice of a spatial hypersurface containing that event \cite{Kuchar,Isham2b}.
\item The solution (\ref{e7a}) of the constraints typically leads to a Hamiltonian
that is, at best, horribly unwieldy.  Consider two popular choices: 
$T_{\hbox{\scriptsize int}} = \sqrt{q}$ (``intrinsic time'') and 
$T_{\hbox{\scriptsize ext}} = q_{ij}\pi^{ij}/\sqrt{q}$ (``extrinsic time'').   
If we choose $T_{\hbox{\scriptsize int}}$ as our time function, we must solve 
the Hamiltonian constraint for its conjugate variable, which is essentially 
$T_{\hbox{\scriptsize ext}}$.   Since $\cal H$ is quadratic in momenta, (\ref{e7a}) 
yields a Hamiltonian $h_T$ proportional to a square root of a functional 
differential operator \cite{Blyth}.  But the operator inside the square root need 
not be positive, and even when it is, the only way to define its square root,
through a spectral decomposition, is highly nonlocal.  

If we instead choose $T_{\hbox{\scriptsize ext}}$ as our time function, we must 
solve the Hamiltonian constraint for $T_{\hbox{\scriptsize int}}$.  But the 
constraint is then a complicated elliptic differential equation---see section 
\ref{seceb}---and $h_T$ can be given only implicitly.  

In either case, the Hamiltonian is plagued with severe operator ordering 
ambiguities, and it is not obvious that it can be defined as a Hermitian operator 
at all.  These problems are not just ``technical'': the difficulty of defining products 
of noncommuting operators at a single point underlies the divergences in ordinary 
quantum field theory, and the renormalization problems of section \ref{secda} are 
likely to reappear here.
\item Perhaps most serious, it is not clear that different choices of a classical time 
function $T$ lead to equivalent quantum theories.  Classically, of course, the choice 
is irrelevant: this is simply an expression of general covariance.  But quantum
mechanically, even in as simple a model as a scalar field in flat spacetime, the 
analogous procedure leads to theories that depend on the choice of time slicing, 
and the theories corresponding to different slicings can be unitarily inequivalent 
\cite{Varadarajan}.  This possible loss of general covariance will reappear below
when we discuss reduced phase space quantization.
\end{itemize}

\subsection{Reduced phase space methods \label{seceb}}

We next turn to reduced phase space quantization (``constrain, then quantize'').
This approach has traditionally been less popular than Dirac quantization, in part 
because the first step---``solve the classical constraints''---is so difficult.  
Nonetheless, considerable work has gone into studying models and looking for 
potential pitfalls.

Like Dirac quantization, reduced phase space quantization requires a series of steps:
\begin{enumerate}
\item Solve the classical constraints (\ref{c16})--(\ref{c17}).  The solutions will
lie on a subspace of the full phase space, the ``constraint surface'' $\bar\Gamma$.
\item The group $\cal G$ of symmetries generated by the constraints, the surface 
deformation group of section \ref{seccb}, acts on the constraint surface.  Factor out this 
action (or gauge-fix the symmetries) to obtain the ``physical'' or ``reduced'' phase 
space ${\hat\Gamma} = {\bar\Gamma}/{\cal G}$.
\item The reduced phase space inherits a symplectic structure (that is, a set of
Poisson brackets) from the brackets (\ref{c14}).  One can also write the action 
(\ref{c15}) in terms of the reduced phase space variables.  The reduced phase 
space action will no longer have constraints, but will typically have a nontrivial 
Hamiltonian.  Quantize this system.
\item Depending on the topology of $\Sigma$, there may still be discrete symmetries 
coming from ``large'' diffeomorphisms.  Impose these as symmetries of the wave 
functions.
\item Figure out how to interpret the resulting wave functions and operators.
\end{enumerate}
Once again, none of these steps is particularly easy:

{\bf Steps 1 and 2}: Symplectic reduction, the process of eliminating the 
constraints, is understood well in principle \cite{Fischer2,Isenberg,Isenberg2}.  
For spatially compact topologies, we know that $\hat\Gamma$ is a stratified 
manifold, with lower-dimensional strata corresponding to spacetimes with 
symmetries; that it has four degrees of freedom per point (twelve degrees of 
freedom in $\{q_{ij},\pi^{ij}\}$ minus four constraints and four symmetries); 
and that it does, indeed, inherit a symplectic structure from the brackets (\ref{c14}) 
\cite{Isenberg3}.  On the other hand, finding a {\em useful\/} parametrization of 
$\hat\Gamma$ is extremely difficult: solving the constraints is certainly not trivial!

Interesting progress in this direction has recently been made by Fischer and 
Moncrief \cite{Fischer3,Fischer4}.  Starting from an old idea 
of York's \cite{York}, they conformally rescale the metric and decompose the 
momentum into irreducible pieces,
\begin{eqnarray}
q_{ij} &=& \phi^4{\bar q}_{ij}  \\
\pi^{ij} &=& {1\over16\pi G}\left( \phi^{-4}p^{ij} 
  - {2\over3}\mathrm{Tr}K\phi^2{\bar q}^{ij}\sqrt{\bar q}
  + (LY)^{ij} \sqrt{q} \right) \quad
  \hbox{with\ \  ${\bar q}_{ij}p^{ij} = 0$, ${\bar\nabla}_j p^{ij} = 0$}  \nonumber
\label{e8}
\end{eqnarray}
where $(LY)^{ij} = \nabla^iY^j + \nabla^jY^i - {2\over3}q^{ij}\nabla_kY^k$ is
the traceless symmetric derivative and $p^{ij}$ is the ``transverse traceless''
component of the momentum.  They then fix the time slicing and the conformal 
factor by requiring that
\begin{eqnarray}
\mathrm{Tr}K &=& -8\pi G {q_{ij}\pi^{ij}\over\sqrt{q}} = -t \nonumber\\
{}^{(3)}\!R[{\bar q}] &=& k \quad \hbox{with $k=0,\pm1$} .
\label{e9}
\end{eqnarray}
The $\mathrm{Tr}K$ equation is York's ``extrinsic time'' gauge; it tells us to 
foliate spacetime by spacelike hypersurfaces of constant mean extrinsic curvature.  
Such a time slicing often exists \cite{Tipler2}, and is conjectured to always exist 
for spatially compact spacetimes; it amounts physically to using the rate of expansion 
of the Universe as time.  The restriction on ${}^{(3)}\!R$ means that ${\bar q}_{ij}$ 
is a ``Yamabe metric.''  The existence and (near) uniqueness of such a conformal 
rescaling follows from Schoen's proof of the Yamabe conjecture \cite{Schoen}; the 
value of $k$, the ``Yamabe type'' of the spacetime, is determined entirely by the 
topology of $\Sigma$.  

The momentum constraints now reduce to equations for $Y$, whose solution
requires that $(LY)^{ij}=0$.  The vector $Y$ thus drops out of the momentum
(\ref{e8}).  For spatial topologies of Yamabe type $-1$, a case that includes a 
wide range of interesting topologies, the Hamiltonian constraint becomes an 
elliptic partial differential equation,
\beq
{\bar\Delta}\phi - {1\over8}\phi + {1\over12}\left( t^2 - 3\Lambda\right)\phi^5
 - {1\over8}{{\bar q}_{ij}{\bar q}_{kl}p^{ik}p^{jl}\over{\bar q}^2}\phi^{-7} 
 = 0, 
\label{e10}
\eeq
which has a unique solution $\phi = \phi({\bar q},p)$.  The reduced phase space 
is thus parametrized by pairs $\{ {\bar q}_{ij},p^{ij}\}$ of Yamabe metrics and 
transverse traceless momenta.   

{\bf Step 3}:  The phrase ``quantize this system'' can hide a multitude of sins.  First,
we must choose a polarization.  In the Fischer-Moncrief approach, it can be shown 
that the reduced phase space is a cotangent bundle over ``conformal superspace,'' 
the space of diffeomorphism classes of Yamabe metrics $\bar q$, and this provides 
a natural guess for a polarization.  But experience with (2+1)-dimensional gravity 
tells us that it is by no means the unique (or even the uniquely physically reasonable) 
choice \cite{Carlip2}.

Next, we must confront a major problem: the Hamiltonian in this formalism is a 
complicated and highly nonlocal function of the reduced phase space degrees of
freedom.  In the Fischer-Moncrief approach, for example, the ADM action reduces
to
\beq
I = \left( {1\over16\pi G}\right)^2 \int dt\int d^3x \left[ p^{ij}\partial_t{\bar q}_{ij}
 - {4\over3}\sqrt{\bar q}\phi^6 \right] ,
\label{e11}
\eeq
with $\phi$ implicitly defined through (\ref{e10}).  Even if we could solve (\ref{e10}),
we would face horrible ordering ambiguities in any attempt to make the ``Hamiltonian'' 
term in (\ref{e11}) into an operator, with a possible reappearance of ultraviolet
divergences.  

Such nonlocality is inherent in reduced phase space methods, in which differential
equations are used to eliminate degrees of freedom.  Electromagnetism, for example, 
has a constraint given by Gauss's law, $\nabla\cdot{\bf E} = 4\pi\rho$.   This
constraint can be solved by splitting the electric field into a ``physical'' transverse 
component ${\bf E}^\perp$ and a longitudinal component ${\bf E}^\parallel = 
\nabla \int d^3y G(x-y)\rho(y)$.  ${\bf E}^\perp$ is then a field in the reduced 
phase space, but ${\bf E}^\parallel$ also contributes to the Hamiltonian, giving 
a nonlocal piece that can be recognized as the Coulomb energy.  For electromagnetism, 
this nonlocal term is harmless, since it is independent of the ``physical'' degrees of 
freedom.  But gravitational energy itself gravitates, and no such simplification can be 
expected in quantum gravity.

{\bf Step 4}:  The momentum constraints generate infinitesimal spatial
diffeomorphisms, and thus account for any diffeomorphism that can be built 
up from infinitesimal transformations.  But for many spatial topologies, the
group of diffeomorphisms is not connected; there are additional ``large''
diffeomorphisms that cannot be generated by the constraints.  The group
of large diffeomorphisms, sometimes called the mapping class group, is  
known to play an important role in questions ranging from the spin of geons 
\cite{Friedman,Carlip7} to the existence of theta vacua \cite{Isham6}.  We 
know from the example of quantum gravity in 2+1 dimensions, where steps 1--3 
are comparatively simple, that the treatment of such large diffeomorphisms is 
potentially quite difficult \cite{Carlip8}; the mapping class group can have a 
very complicated action on the reduced phase space, which may not easily
project down to an action on the configuration space.

{\bf Step 5}:  Let us suppose we have solved all of the problems of formally
constructing a reduced phase space quantization.  Some of the difficulties of
Dirac quantization will have disappeared: there are no longer any constraints;
wave functions are just functions (or sections of a bundle)  on a reduced 
configuration space with an ordinary inner product; observables are functions 
and derivatives on that space; and we now have an ordinary, albeit complicated, 
Hamiltonian.  But is this the quantum theory of gravity we want?

In passing to the reduced phase space, we froze many degrees of freedom
classically, forbidding their quantum fluctuations.  It is not clear that this 
is the right thing to do.  In a path integral, for instance, {\em all\/} histories are 
at least nominally included.   By treating only a subset of ``physical'' fields
quantum mechanically, we may be excluding physically important quantum
effects.

Even more worrisome is the treatment of time \cite{Kuchar,Isham2b}.  In the 
Fischer-Moncrief reduction we fixed time to be $\mathrm{Tr}K$, and can easily
define quantum states only on $\mathrm{Tr}K = \mathit{const.}$ slices.  Other 
methods of reduction involve different choices of a time slicing, but {\em some\/} 
choice must be made.  Even if we are willing to accept this restriction on initial 
and final states, it is by no means clear that different time slicings in intermediate 
regions will give the same transition amplitudes---reduced phase space quantization 
may not preserve general covariance.  The results of Torre and Varadarajan again 
tell a cautionary tale \cite{Varadarajan}: even for a free scalar field in flat spacetime 
with fixed initial and final spatial hypersurfaces, different intermediate time slicings 
lead to inequivalent quantum theories.  Of course, many problems can be translated 
into operator ordering ambiguities, and it could be that a proper choice of ordering 
restores covariance \cite{Cosgrove}, but this remains largely a hope.

\subsection{Covariant canonical quantization \label{secec}}

From a relativist's point of view, the canonical methods of this section share
a fundamental weakness: all require a split of spacetime into space and time,
violating at least the spirit of general covariance.  The covariant methods of
the preceding section share a different weakness: they are perturbative, and 
require a ``nice'' classical background.  One approach, covariant canonical 
quantization, avoids both of these weaknesses, though only at the expense of 
immense and perhaps unsolvable technical difficulty.

The starting point for covariant canonical quantization is the observation that 
the phase space of a well-behaved classical theory is isomorphic to the space of 
classical solutions, or ``histories'' \cite{Ashtekar5,Ashtekar6,Crnkovic,Lee2,Wald4}.  
To see this, let $\cal C$ be an arbitrary (but fixed) Cauchy surface.  Then a point 
in phase space gives initial data on $\cal C$, which determine a unique solution,
while conversely, a classical solution restricted to $\cal C$ determines a point in 
phase space.  Wald et al.\ have found an elegant description of the resulting symplectic 
structure on the space of histories \cite{Lee2,Wald4}.  For the first order action 
(\ref{c28a}), the symplectic form (strictly speaking, the presymplectic form---we 
must still factor out gauge invariances) is especially simple:
\beq
\Omega[\delta_1 e,\delta_2\omega] = {1\over32\pi G}\int_{\cal C}
  \epsilon_{IJKL}e^I\wedge\delta_1 e^I\wedge\delta_2\omega^{KL} ,
\label{e12}
\eeq
where $\cal C$ is any Cauchy surface.  One should view $\delta e$ and 
$\delta\omega$ as ``vectors on the space of histories''; $\Omega$ is then an 
antisymmetric two-form on this space, and defines a symplectic structure.  If 
we parametrize the space of classical solutions by variables $(q_\alpha,p_\beta)$, 
(\ref{e12}) is, schematically, $\Omega[p,q] = \Omega^{\alpha\beta}
q_\alpha p_\beta$, and the corresponding Poisson brackets are
\beq
\{q_\alpha,p_\beta\} = \Omega^{-1}{}_{\alpha\beta} .
\label{e13}
\eeq
The idea of covariant canonical quantization is to ``quantize these brackets.''

Since the parameters $(q_\alpha,p_\beta)$ describe entire classical solutions,
they are automatically diffeomorphism invariant.  In particular, there is no need to 
choose a time slicing; the symplectic form (\ref{e12}) is independent of $\cal C$.  
Moreover, since we began with classical solutions, the constraints are trivially 
satisfied.  The $(q_\alpha,p_\beta)$ are thus observables in the strong sense:   they 
are diffeomorphism-invariant quantities that have vanishing Poisson brackets 
with the constraints.  Kucha{\v r} has coined the term ``perennials'' for such 
observables to emphasize their time independence.

I have, of course, concealed a multitude of problems.  Apart from the difficulty of 
finding a complete set of perennials $(q_\alpha,p_\beta)$---essentially, finding 
the general solution of the classical field equations---I have assumed a polarization, 
a splitting of phase space into ``positions'' and ``momenta.''  In 2+1 dimensions, this 
turns out to be easy \cite{Carlip2}: the space of histories is a cotangent bundle, with 
a base space parametrized by spin connections.  But life is more complicated in 
3+1 dimensions, and there may not be a preferred choice.

The big problem, though, lies in determining physically interesting observables.  
Wave functions in covariant canonical quantization are functions on the space of 
spacetimes, and are thus inherently time independent.  We obtain a sort of Heisenberg 
picture, in which time dependence must reside in the operators.  But the ``perennials''
$(q_\alpha,p_\beta)$ are also time independent, and cannot in themselves describe
evolution.

One proposal to solve this problem is due to Rovelli \cite{Rovelli4,Rovelli5}: 
observables should be ``evolving constants of motion,'' one-parameter families 
${\cal O}_t$ of operators on the space of histories that agree classically with 
time-dependent observables at time $t$.   Consider, for example, the family of 
classical functions $V_t[g] =$ ``the volume of the spacelike hypersurface of mean 
curvature $\mathrm{Tr}K = -t$ in the spacetime characterized by the metric 
$g$.''   For each label $t$, $V_t$ is a function on the space of solutions, and in 
principle can be expressed in terms of  $(q_\alpha,p_\beta)$.  Again in principle, 
we can turn such a function into an operator.  In practice, we quickly encounter the 
same problems of nonlocality and ordering ambiguities that plague other methods.  
These can lead to disaster: ``bad'' choices of observables can lead to inconsistent 
time orderings and different predictions for the same history \cite{Hartle4}.  

A related problem has been found by H{\'a}j{\'\i}{\v c}ek: different initial choices 
of the parameters $(q_\alpha,p_\beta)$ can lead to unitarily inequivalent quantum 
theories \cite{Hajicek3}.   The problem, slightly oversimplified, is that spacetime 
coordinates should not be observable---the theory is, after all, covariant---but 
metric-dependent classical coordinate transformations can mix coordinates and 
observables.  Note that such metric-dependent coordinate transformations are not 
easy to avoid: in even the simple example of the transformation from Schwarzschild 
to Kruskal coordinates, the transformation depends on the (observable) mass.  
The issue can be rephrased in terms of the ``reconstruction problem'' of section 
\ref{secb}, as the statement that the ``metric operator,'' the ``position operator,'' 
and similar quantities built from the fundamental variables of covariant canonical 
quantization are not uniquely defined.

Because of the immense technical problems, fairly little progress has been made in
covariant canonical quantization except in simple models.  In (2+1)-dimensional 
general relativity, the program can be carried out in full for a few spatial topologies 
\cite{Carlip,Carlip2}, and one can find operators that reproduce the results of 
reduced phase space quantization with the York time slicing.  Attempts to quantize 
asymptotic ``radiative'' gravitational degrees of freedom use similar methods 
\cite{Ashtekar7}.  It is an open question whether these results can be extended, 
perhaps perturbatively, to a full quantum theory of gravity.

\section{Loop quantum gravity and ``quantum geometry'' \label{secf}}
\setcounter{footnote}{0}

In 1986, building on an observation of Sen \cite{Sen}, Ashtekar suggested that the
right variables for quantum gravity were the self-dual connection variables
of sections \ref{seccc}--\ref{seccd}, and pointed out  the usefulness of the 
connection representation \cite{Ashtekar2}.  Those suggestions have blossomed
into a major program, ``loop quantum gravity'' or ``quantum geometry.''  Much 
of the work in this area has been based on Dirac quantization of the constraints 
(\ref{c41}), though there have been recent advances in the use of covariant 
``spin foam'' methods \cite{Baez}.  Since a thorough review of this subject already 
exists \cite{Rovelli} (see also \cite{Ezawa,Baez2}), my treatment will be sketchy, 
emphasizing the relationships to other approaches to quantum gravity.

\subsection{Kinematics and spin networks \label{secfa}}

Loop quantum gravity is based on Dirac quantization of the set of constraints 
(\ref{c41}).  The program is greatly simplified if one uses the real connection 
formulation described at the end of section \ref{seccd}.  Accordingly, let us allow 
an arbitrary Immirzi parameter $\gamma$, with the general Hamiltonian constraint 
(\ref{c49}).  We begin in the connection representation; that is, we start with a space 
of functionals $\Psi[A]$ of the connection.  By (\ref{c48}), this means 
\beq
{\tilde E}^i{}_{\hat I} = -8\pi\gamma G{\delta\ \over\delta A_i{}^{\hat I}} .
\label{f1}
\eeq
One of the important achievements of loop quantization is its ability to make 
sense of expressions of this sort as well-defined operators on a genuine Hilbert 
space.

We start with the Gauss law constraint ${\cal G}^{\hat I}=0$.  As in ordinary
Yang-Mills theory, ${\cal G}^{\hat I}$ generates ($\hbox{SO}(3)$ or 
$\hbox{SU}(2)$) gauge transformations, and the constraint requires that states 
be gauge-invariant functionals of the connection $A$.  Fortunately, we know how 
to find such functionals.  For any curve $\gamma: [0,1]\rightarrow\Sigma$, consider 
the holonomy
\beq
U_\gamma(s_1,s_2) =  
P\exp\left\{ -\int_{s_1}^{s_2} ds {dx^i(s)\over ds}A_i{}^{\hat I}\tau_{\hat I} \right\} ,
\label{f2}
\eeq
where $P$ denotes path ordering.  Then for a closed curve, the ``Wilson loop''
$\mathrm{Tr}\,U_\gamma(0,1)$ is gauge invariant.  More generally, let $\Gamma$ 
be a graph, and define a ``coloring'' as follows: 
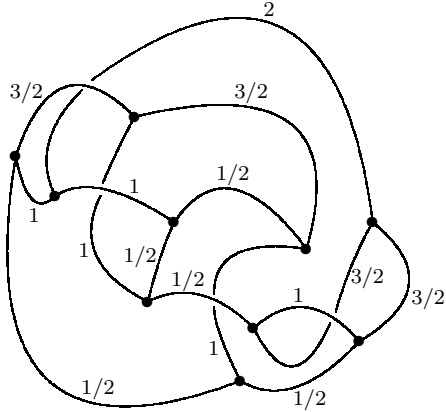
\begin{figure}
\begin{picture}(150,150)(-180,-80)

\put(0,0){\circle*{4}}
\qbezier(0,0)(20,30)(50,-10)
\put(16,16){\scriptsize $1/2$}

\put(50,-10){\circle*{4}}
\qbezier(0,0)(-30,20)(-45,10)
\put(-17,11){\scriptsize $1$} 

\put(-45,10){\circle*{4}}
\qbezier(0,0)(-5,-10)(-10,-30)
\put(-19,-15){\scriptsize $1/2$}

\put(-10,-30){\circle*{4}}
\qbezier(-10,-30)(10,-20)(30,-40)
\put(-1,-24){\scriptsize $1/2$}

\put(30,-40){\circle*{4}}
\qbezier(-10,-30)(-40,-15)(-28,10)
\put(-36,-13){\scriptsize $1$}
\qbezier(-27,15)(-22,27)(-15,40)
\put(-15,40){\circle*{4}}
\qbezier(-15,40)(70,60)(50,-10)
\put(23,47){\scriptsize $3/2$}

\qbezier(-15,40)(-45,70)(-60,25)
\put(-62,47){\scriptsize $3/2$}

\put(-60,25){\circle*{4}}
\qbezier(-60,25)(-55,0)(-45,10)
\put(-55,0){\scriptsize $1$}

\qbezier(50,-10)(15,-5)(15,-27)
\qbezier(15,-32)(15,-40)(25,-60)
\put(13,-50){\scriptsize $1$}
\put(25,-60){\circle*{4}}

\qbezier(-60,25)(-80,-100)(25,-60)
\put(-35,-65){\scriptsize $1/2$}

\qbezier(30,-40)(50,-22)(70,-45)
\put(45,-30){\scriptsize $1$}
\put(70,-45){\circle*{4}}

\qbezier(25,-60)(45,-72)(70,-45)
\put(45,-69){\scriptsize $1/2$}

\qbezier(-45,10)(-55,30)(-35,50)
\qbezier(-31,54)(60,120)(75,0)
%\put(0,75){\scriptsize $2$}
\put(34,78){\scriptsize $2$}
\put(75,0){\circle*{4}}

\qbezier(30,-40)(45,-70)(59,-38)
\qbezier(61,-34)(66,-16)(75,0)
\put(67,-23){\scriptsize $3/2$}

\qbezier(70,-45)(105,-24)(75,0)
\put(90,-31){\scriptsize $3/2$}

\end{picture}
\caption{A trivalent spin network.  The intertwiners at each vertex are
ordinary Clebsch-Gordan coefficients. \label{fig3}}
\end{figure}

\begin{enumerate}
\item to each edge, assign a half-integer $s_i$ labeling an irreducible representation 
of $\hbox{SU}(2)$;
\item to each vertex at which spins $s_1,\dots,s_n$ meet, assign an intertwiner, that 
is, an invariant tensor $v_\alpha$ in the tensor product of the representations 
$s_1,\dots,s_n$.
\end{enumerate}
Such a colored graph $S = \{\Gamma, s_i, v_\alpha\}$, introduced by Penrose
\cite{Penrose3}, is called a spin network.\footnote{For an introduction to spin 
networks and spin network technology, see \cite{Ezawa} and \cite{Major}.}  Each 
spin network determines a gauge-invariant functional of the connection, evaluated 
by a simple algorithm: for each edge labeled by spin $s_i$, find the holonomy of $A$ 
in the representation $s_i$, and use the intertwiner at each vertex to combine the 
holonomies into an invariant.  This functional can be viewed as a state 
$\psi_S[A]$ in the connection representation.

In contrast to the mathematical imprecision of the metric representation of 
section \ref{secea}, Ashtekar, Baez, Lewandowski, and others have shown how 
to define an inner product on the space of gauge-invariant functionals of the 
connection and make it into an honest Hilbert space.  Equivalently, one can 
do integral and differential calculus on the space $\overline{{\cal A}/{\cal G}}$
of gauge-equivalence classes of generalized connections \cite{Baez2}.   The
spin network states form an orthogonal basis for this Hilbert space; any state can 
be written as a superposition,
\beq
|\psi\rangle = \sum_S \langle S|\psi\rangle |S\rangle
\label{f3}
\eeq
where the coefficient $\langle S|\psi\rangle$ is the inner product of  $\psi[A]$ with 
the basis state $\psi_S[A]$,
\beq
\langle S|\psi\rangle = \int _{\overline{{\cal A}/{\cal G}}}
  d\mu[A] \psi_S^*[A]\psi[A] .
\label{f4}
\eeq
The coefficients $\langle S|\psi\rangle$ thus completely determine a state,
giving a ``spin network representation.''   Such representations were originally 
defined for Wilson loops \cite{RovelliSmolin}; hence the term ``loop quantum 
gravity.''

We can next define a natural set of gauge-invariant ``loop operators,''  
\begin{eqnarray}
&&{\cal T}[\gamma] = - \mathrm{Tr}\, U_\gamma(0,1) \nonumber\\
&&{\cal T}^i[\gamma](s) = - \mathrm{Tr}\, [U_\gamma(s,s){\tilde E}^i(s)] 
  \nonumber\\
&&{\cal T}^{i_1\dots i_N}[\gamma](s_1,\dots,s_N) = - \mathrm{Tr}\, 
  [U_\gamma(s_1,s_N){\tilde E}^{i_N}(s_N)\dots
  U_\gamma(s_2,s_1){\tilde E}^{i_1}(s_1)] .
\label{f5}
\end{eqnarray}
Using (\ref{f1}), we obtain well-defined operators on the spin network states, 
with a calculable commutator algebra.  In particular, the algebra of the lowest 
order operators, ${\cal T}$ and ${\cal T}^i$, gives a faithful representation 
of their classical Poisson algebra.  

Other more ``geometric'' operators can also be defined, notably an area operator 
${\bf A}$ and a volume operator ${\bf V}$.  Such operators contain products 
of functional derivatives at a point---${\bf A}$, for instance, depends on 
$({\tilde E})^2$, or on ${\cal T}^{ij}(s_1,s_1)$---and  must be regularized.  
Thanks to the discrete nature of spin network states, though, they behave better 
than one might expect, and their complete spectra can be computed.  By 
(\ref{f1}), the results depend on the Immirzi parameter.  The area operator, for
example, has eigenvalues of the form
\beq
A = 8\pi\gamma G \sum_i\sqrt{j_i(j_i + 1)} ,
\label{f6}
\eeq
where $\{j_i\}$ are a set of integers and half-integers; see \cite{Upadhya} for a 
simple computation.  The $\gamma$ dependence remains mysterious, and may 
represent a genuine quantization ambiguity; it will be important in the discussion
of black hole entropy in section \ref{secib}.

We next turn to the momentum constraints ${\cal V}_i=0$.  As in the ADM
picture, these constraints generate spatial diffeomorphisms, and their action on 
spin network states is essentially to drag the graph $\Gamma$.  To solve the 
constraints, we simply ``forget'' the embedding of $\Gamma$ in $\Sigma$, and
consider only its topologically invariant knotting and linking properties.  States
thus depend only on ``knots,'' or more precisely ``s-knots,'' equivalence classes
of spin networks.  This can all be made mathematically rigorous, and we can
again define a genuine Hilbert space $H^{\hbox{\scriptsize (Diff)}}$ with a 
calculable inner product \cite{Ashtekar8}.  

The area and volume operators are not invariant under the transformations
generated by ${\cal V}_i$.  If $\sigma\subset\Sigma$ is a fixed surface, for 
example, ${\cal V}_i$ will act on the phase space variables in ${\bf A}(\sigma)$ 
but not on $\sigma$ itself.  ${\bf A}$ and ${\bf V}$ are thus not defined on  
$H^{\hbox{\scriptsize (Diff)}}$.  If a surface $\sigma$ or volume $v$ is defined 
by {\em physical\/} variables, however---``the surface on which the scalar field 
$\varphi$ vanishes,'' for instance---then the constraints will act on $\sigma$ 
or $v$ as well, and ${\bf A}(\sigma)$ and ${\bf V}(v)$ will be good operators 
on $H^{\hbox{\scriptsize (Diff)}}$.  This is an expression of the ``reconstruction
problem'' of section \ref{secb}: one no longer has a background spacetime to label 
geometric objects, but must characterize them intrinsically in a way that respects 
the symmetries and the dynamics.

\subsection{Dynamics \label{secfb}}

We have now completed steps 1--3 and half of step 4 of section \ref{secea}.  
So far, we are in good shape: we have well-defined operators acting on a 
well-defined ``kinematical'' Hilbert space, and have even obtained some
concrete predictions like the quantization of area.  It remains to complete the 
program by solving the Hamiltonian constraint, forming a true physical Hilbert 
space, and interpreting the resulting theory.

An original rationale for Ashtekar variables was that they simplified the 
Hamiltonian constraint.  But this simplification requires the choice $\gamma=i$, 
and thus a complex connection and problematic reality conditions.  Spin network 
technology, in contrast, is based on real $\hbox{SU}(2)$ connections.   There 
are three basic strategies for dealing with this mismatch:
\begin{enumerate}
\item We can try to relate real and complex connections via a ``Wick rotation''
\cite{Thiemann,Ashtekary}.  For metrics with {\em Riemannian\/} signature, 
a real Immirzi parameter $\gamma=1$ leads to a simple Hamiltonian constraint 
of the form (\ref{c41}).  We certainly cannot simply work with Riemannian 
metrics and then analytically continue in time, but we can find an operator 
${\hat T}$ that transforms between the ``Lorentzian'' Hamiltonian constraint
 and the simpler ``Riemannian'' one.  It is formally true that
\beq
{\cal S}_{\hbox{\scriptsize\it Lor}}  
  = e^{-i{\hat T}}{\cal S}_{\hbox{\scriptsize\it Riem}} e^{i{\hat T}} \qquad
\hbox{with}\quad
{\hat T} = {i\over16G}\int_\Sigma d^3x K_i{}^{\hat I}{\tilde E}^i{}_{\hat I} 
\label{f9}
\eeq
where ${\cal S}_{\hbox{\scriptsize\it Lor}}$ is the complicated constraint 
(\ref{c49}) for $\gamma=1$ and ${\cal S}_{\hbox{\scriptsize\it Riem}}$ is
the simpler ``Riemannian'' equivalent.  If we could really define this operation 
on spin network states, we could solve the simpler ``Wick rotated constraint'' 
${\cal S}_{\hbox{\scriptsize\it Riem}}|\psi_ {\hbox{\scriptsize\it Riem}}\rangle
= 0$ and then ``rotate back'' the wave function to $|\psi_{\hbox{\scriptsize\it Lor}}
\rangle = \exp\{-i{\hat T}\}|\psi_{\hbox{\scriptsize\it Riem}}\rangle$.

This is not easy:  $K_i{}^{\hat I}$ is defined by (\ref{c47}), and is thus a 
complicated functional of the triad.   Thiemann has suggested a clever trick for 
defining this and similar operators in terms of commutators 
\cite{Thiemann,Thiemann2}.  Classically, it is easily checked 
that
\beq
{\cal S}_{\hbox{\scriptsize\it Riem}} = {1\over4\pi G}\sqrt{q}\epsilon^{ijk}
   F_{ij{\hat K}}\left\{ A_k{}^{\hat K},{\bf V}(\Sigma)\right\}  
\label{f10}
\eeq
where ${\bf V}(\Sigma) = \int_\Sigma d^3x\sqrt{q}$ is the spatial volume.   
Similarly, $T$ can be expressed in terms of the Poisson bracket of 
${\cal S}_{\hbox{\scriptsize\it Riem}}$ with ${\bf V}$.  Thiemann proposes 
that the corresponding operators be defined on spin network states by replacing 
the Poisson brackets with commutators and using the established regularization 
of the volume operator.  
\item We can try to make sense of the Hamiltonian constraint
${\cal S}_{\hbox{\scriptsize\it Lor}}$ directly.  Again, the main proposal is to 
use Poisson brackets like (\ref{f10}) to define the action of the constraint
\cite{Thiemann2}.  This can be done \cite{Borissov}, and gives a ``combinatorial'' 
action in which the constraint adds lines and vertices to a spin network in a 
prescribed fashion.  Unfortunately, this approach has a number of problems 
\cite{Smolin2,Gambini2}.  It is, in a sense, ``too local'':  the constraint
acts independently on separate regions of a typical spin network, and solutions
lack long range correlations needed for a good classical limit.  Moreover, 
the commutators of the constraints fail to reproduce the surface deformation 
algebra (\ref{c23}), essentially because the regularization sets the 
inverse spatial metric $q^{ij}$ on the right-hand side of (\ref{c23}) to zero.  
These difficulties are already present in the ``Riemannian'' formulation, making 
the Wick rotation method problematic as well.

Recently, some progress has been made in an alternative approach to the
Hamiltonian constraint \cite{Gambini,Bartolo}, based on a different set of
states constructed from generalized knot invariants (Vassiliev invariants).  But 
here, too, the commutators of the constraints fail to reproduce the surface 
deformation algebra, and the regularized inverse spatial metric vanishes.  This 
is not {\em necessarily\/} a problem, since $q^{ij}$ is not an observable---this 
is again a reflection of the ``reconstruction problem''---but it is a cause for worry.
\item We can return to a four-dimensional picture and try to formulate a path
integral approach that enforces the Hamiltonian constraint.  The canonical
formulation of loop quantum gravity describes states in terms of spin networks,
so it is natural to try to describe the full four-dimensional theory in terms of
evolving spin networks, or ``spin foams'' (see \cite{Baez,Baez3} for reviews).  
The simplest spin foam is just the world sheet of a spin network, that is, a spin
network in which the vertices are stretched out in the temporal direction to 
form lines and the edges are stretched to form surfaces.  More complicated spin 
foams allow transitions between spin networks, as illustrated for instance in 
figure \ref{fig4}.
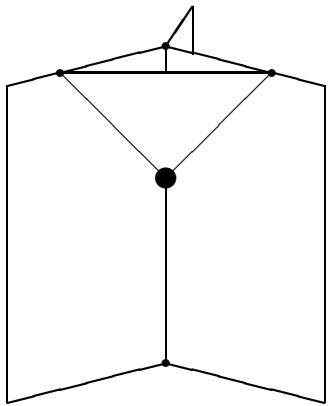
\begin{figure}
\begin{picture}(150,150)(-180,-40)

\put(-30,-40){\line(0,1){120}}
\put(30,-25){\line(0,1){70}}
\put(30,45){\circle*{8}}
\put(30,45){\line(-1,1){40}}
\put(30,45){\line(1,1){40}}
\put(90,-40){\line(0,1){120}}
\put(30,85){\line(0,1){10}}
\put(40,110){\line(0,-1){18}}
\thicklines
\put(-30,-40){\line(4,1){60}}
\put(30,-25){\line(4,-1){60}}
\put(-30,80){\line(4,1){60}}
\put(90,80){\line(-4,1){60}}
\put(30,95){\line(2,3){10}}
\put(-10,85){\line(1,0){80}}
\put(30,-25){\circle*{3}}
\put(30,95){\circle*{3}}
\put(-10,85){\circle*{3}}
\put(70,85){\circle*{3}}

\end{picture}
\caption{A  simple spin foam vertex, representing the creation of a new line
and two new vertices in a spin network. \label{fig4}}
\end{figure}

The impetus for this line of research comes from several directions.  First, one
can formally impose the Hamiltonian constraint by looking at amplitudes 
$\langle\psi | \exp\{it{\hat H}\} |\phi\rangle$, where $\hat H$ is an integrated 
version of the Hamiltonian constraint, and then integrating over $t$.  Reisenberger 
and Rovelli have shown that given some fairly general assumptions about $\hat H$, 
such amplitudes can be expressed in terms of spin foams \cite{Reisenberger2}, 
which act as ``Feynman diagrams'' in the expansion of the propagator.  Spin foams 
also appear in certain lattice formulations of gravity.  Finally, spin foams are known 
to give correct amplitudes and partition functions for topological $BF$ theories, 
and gravity, as we saw in section \ref{secce}, can be viewed as a constrained $BF$ 
theory.  This last perspective suggests interesting interpretation of spin foams as 
``quantum four-geometries'' \cite{Baez3}.

Existing spin foam models depend on a triangulation of the underlying  spacetime:
each vertex of the spin foam corresponds to a four-simplex of the manifold,
each edge to a tetrahedron, and each face to a triangle.  This causes no problem 
for unconstrained $BF$ theory, for which amplitudes are independent of the 
triangulation, essentially because $BF$ theory has no local degrees of freedom.  
But a fixed triangulation in quantum gravity truncates the degrees of freedom, and 
is clearly inappropriate.  It may be possible to carry the ``Feynman diagram'' analogy 
further, and treat spin foams as diagrams that must be summed to include all 
triangulations.  It is formally possible to write a field theory for which spin foams 
really are Feynman diagrams \cite{Reisenberger3}, though the deeper significance 
of such a theory remains unclear.  

Most spin foam research has focused on the relatively simple case of Riemannian 
signature.  Recently, though, there have been interesting attempts to define 
Lorentzian spin foams, both as geometrically-motivated generalizations of the 
Riemannian constructions  \cite{Barrett} and from the perspective of causal sets 
\cite{Markopoulou2}.   
\end{enumerate}

Despite the problems in defining the Hamiltonian constraint, some exact solutions
are known \cite{Ezawa}.  If the cosmological constant vanishes, the diffeomorphism
and Hamiltonian constraints are both proportional to $F_{ij}{}^{\hat I}$, so any
gauge-invariant wave function that has its support only on flat connections is
automatically a solution \cite{Blencowe}.  For $\Lambda\ne0$, wave functions
proportional to the exponential of the Chern-Simons action for the connection 
$A$ solve all the constraints \cite{Kodama,Brugmann}.  Such solutions are at best 
a very small piece of the total physical Hilbert space, but the ability to write down 
{\em any\/} exact solution in quantum gravity is something of a breakthrough.

\subsection{Conceptual issues}

Finally, we come to the problem of interpreting loop quantum gravity.  Perhaps 
because of the technical successes of the program, issues of interpretation are brought 
into sharp relief.  For example, loop quantum gravity with Thiemann's regularization 
of the Hamiltonian constraint appears to be a consistent, diffeomorphism-invariant 
quantum theory obtained by quantizing general relativity.  But as noted above, there 
are indications that it is not ``quantum gravity''---that is, it may not reduced to 
general relativity in the classical limit.

The basic problem is the ``reconstruction problem'': how, given a collection 
of states and operators, does one recover a geometrical picture of spacetime?  At 
the Planck scale, the states of loop quantum gravity don't look like three-space: 
the ``geometry'' is restricted to a discrete, one dimensional network.  They 
{\em certainly\/} don't look like spacetime: we chose from the beginning to work 
on a fixed time slice.  This is not necessarily a bad thing---quantum gravity should 
be expected to transform our microscopic picture of spacetime---but the result 
cannot be called gravity until we can understand how the classical picture emerges.

One strength of loop quantum gravity is the existence of ``geometric'' operators
like area and volume, from which one might hope to build up an approximate
classical geometry.  Unfortunately, these operators do not commute with the
constraints, so they fail to take physical states to physical states.  It may be 
possible to define ``evolving constant of motion'' \cite{Rovelli4,Rovelli5} that 
determine invariant information about such geometric quantities, but we do 
not yet know how, except by artificially adding extra fields to pick out surfaces 
\cite{Smolin3,Ashtekar9}.

These problems would be far more tractable if we knew how to find a classical 
limit for loop quantum gravity.   Work in this area includes the construction 
of ``weave states'' approximating fixed classical backgrounds 
\cite{Ashtekar9,Zegwaard,Iwasaki}; an attempt to build spin network states 
from random classical geometries with fixed statistical properties\cite{Bombelli}; 
and efforts to define coherent states \cite{Sahlmann}.   While some progress 
has been made in building states with reasonable semiclassical behavior, it remains 
unclear why these particular states should be the ones relevant to our macroscopic 
world.

\section{String theory}\setcounter{footnote}{0}

I have so far avoided discussing what is arguably the most popular current approach 
to quantum gravity, string theory.  String theory is much more than an attempt
to quantize gravity; it is a proposal for a ``theory of everything,'' in which gravity
would take its place beside the other fundamental interactions.  As such, it is a huge
subject, big enough for textbooks (reference \cite{Polchinski} is the most recent) and 
extensive review articles, including one in this journal \cite{Danielsson}.  I will make
no attempt to review this area; rather, I will focus on a few points at which string
theory is most relevant to the problem of quantizing gravity.\footnote{In 
this section, lower case Roman letters $a$, $b$, \dots from the beginning of the 
alphabet are world sheet indices, ranging from $0$ to $1$.  Spacetime coordinates 
are $X^\mu$; world sheet coordinates are $\sigma^a$ or $\sigma$, $\tau$.}

\subsection{Perturbative string theory \label{secga}}

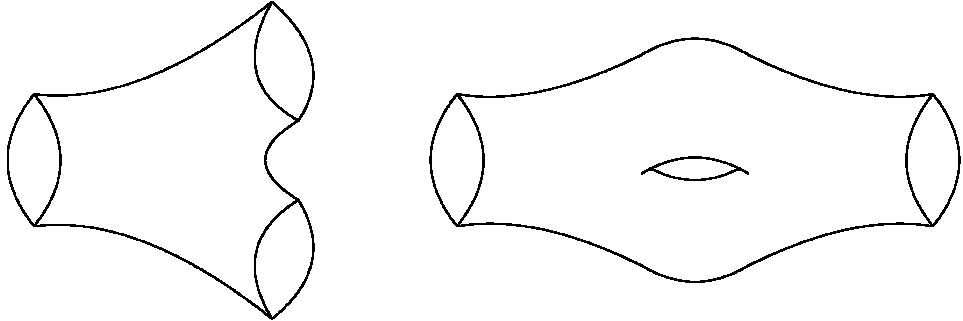
\begin{figure}
\begin{picture}(150,150)(-140,-60)

\qbezier(-80,25)(-100,0)(-80,-25)
\qbezier(-80,25)(-60,0)(-80,-25)
\qbezier(-80,25)(-40,20)(10,60)
\qbezier(-80,-25)(-40,-20)(10,-60)
\qbezier(10,60)(-7,30)(20,15)
\qbezier(10,60)(35,38)(20,15)
\qbezier(10,-60)(-7,-32)(20,-15)
\qbezier(10,-60)(35,-40)(20,-15)
\qbezier(20,15)(-5,0)(20,-15)

\qbezier(80,25)(100,0)(80,-25)
\qbezier(80,25)(60,0)(80,-25)
\qbezier(80,25)(110,20)(150,40)
\qbezier(80,-25)(110,-20)(150,-40)
\qbezier(260,25)(230,20)(190,40)
\qbezier(260,-25)(230,-20)(190,-40)
\qbezier(260,25)(280,0)(260,-25)
\qbezier(260,25)(240,0)(260,-25)
\qbezier(150,40)(170,52)(190,40)
\qbezier(150,-40)(170,-52)(190,-40)
\qbezier(150,-5)(170,7)(190,-5)
\qbezier(153,-3)(170,-12)(187,-3)

\end{picture}
\caption{A closed string vertex and a one-loop process. \label{fig5}}
\end{figure}

The basic premise of perturbative string theory is that the fundamental constituents
of matter are one-dimensional ``strings'' rather than point particles.  Strings may 
be closed (loops) or open (line segments); they interact by splitting and merging, 
as in figure \ref{fig5}.  String theory offers an unexpected solution to the problem 
of nonrenormalizability: by replacing pointlike processes with extended interactions 
that cannot be localized, it avoids quantum field theoretical divergences from the 
start.  This idea is surprising not so much because nonlocality is inherently strange, 
but because it is extraordinarily hard to find a consistent nonlocal theory that respects 
Lorentz invariance, much less diffeomorphism invariance, while preserving causality.

As a string moves through a $D$-dimensional background spacetime, it traces 
out a two-dimensional ``world sheet'' $S$, which can be described by an embedding 
$(\sigma,\tau)\rightarrow X^\mu(\sigma,\tau)$.  The simplest string action, the 
Nambu-Goto action, is just the area of this world sheet.  It is usually easier to use 
an equivalent action, introduced by Polyakov \cite{Polyakov}, 
\beq
I_{\hbox{\scriptsize str}} = {1\over4\pi\alpha'}\int_S d^2\sigma\sqrt{g}
  g^{ab}\partial_aX^\mu\partial_bX^\nu \eta_{\mu\nu} + \lambda\chi
\label{g1}
\eeq
where $g_{ab}$ is an auxiliary world sheet metric, $\lambda$ is a 
constant fixed by the dilaton vacuum expectation value, and $\alpha'$ is a 
coupling constant of the order of the square of the Planck length.  In the last 
term, $\chi$ is the Euler number of $S$,
\beq
\chi = {1\over4\pi}\int_S d^2\sigma\sqrt{g}\,{}^{(2)}\! R
  + {1\over2\pi}\int_{\partial S}ds\,k ,
\label{g1a}
\eeq
where $k$ is the geodesic curvature of the boundary.  The Euler number is a 
topological invariant, and does not affect the classical equations of motion, but it 
controls the string coupling strength.  Indeed, each closed string loop decreases 
$\chi$ by 2, and so contributes a factor of $e^{2\lambda}$ in the path integral
\beq
Z_{\hbox{\scriptsize string}} = \int [dg][dX]\exp\{-I_{\hbox{\scriptsize str}}\} .
\label{g1b}
\eeq
We thus have a ``coupling constant'' $g_c=e^\lambda$ for each closed string 
vertex.  Similarly, each open string vertex comes with a coupling constant 
$g_o=e^{\lambda/2}$.

From the point of view of the two-dimensional world sheet, the action (\ref{g1})
describes a scalar field theory, with a set of fields $\{X^\mu\}$ that happen to be 
interpretable as coordinates of an embedding.  In addition to its obvious world 
sheet diffeomorphism invariance and spacetime Lorentz invariance, (\ref{g1}) 
is invariant under Weyl (``conformal'') transformations,
\beq
g_{ab}\rightarrow e^{2\omega(\sigma,\tau)}g_{ab} ,
\label{g2}
\eeq
so this two-dimensional theory is a conformal field theory.  Preservation of Weyl 
invariance in the quantum theory determines key aspects of string theory.

Canonical quantization of the Polyakov action is relatively straightforward,
although the treatment of diffeomorphism invariance requires some care.
Roughly speaking, $X^\mu$ can be expanded in a Fourier series, which for
a closed string takes the form
\beq
X^\mu(\sigma,\tau) =  x^\mu + 2\alpha'p^\mu\tau 
  + i\sqrt{\alpha'\over2}\sum_{n\ne0}\left[{1\over n}\alpha^\mu_n
  e^{-2in(\tau-\sigma)} + {1\over n}{\tilde\alpha}^\mu_n
  e^{-2in(\tau+\sigma)}\right] ,
\label{g3}
\eeq
and the coefficients $\alpha^\mu_n$ and ${\tilde\alpha}^\mu_n$ are found 
to have commutators
\beq
[\alpha^\mu_m,\alpha^\nu_n] = [{\tilde\alpha}^\mu_m,{\tilde\alpha}^\nu_n]
  = m\eta^{\mu\nu}\delta_{m,-n} , \qquad
[\alpha^\mu_m,{\tilde\alpha}^\nu_n] = 0 .
\label{g4}
\eeq
With a proper treatment of the indefinite metric, these act like an infinite
tower of creation and annihilation operators, and one can build states from
the vacuum by acting with a string of creation operators $\alpha_{-n}$
and ${\tilde\alpha}_{-n}$.  It is not hard to compute the masses of such
states---the zero of energy is fixed by the requirements of Weyl and Lorentz 
invariance---and one finds that for closed strings, the state $\alpha_{-1}^\mu
{\tilde\alpha}_{-1}^\nu|0\rangle$ is massless.  The symmetric traceless part 
of this state is thus a massless, spin two field, and by the arguments of section 
\ref{secca}, must be described by an action that looks like the Einstein action 
plus possible higher order corrections \cite{Scherk}.  String theory thus contains 
``gravitons.''

The action (\ref{g1}) describes a string in a flat Minkowski background.  A
second sign that string theory contains gravity appears if we replace the flat
metric $\eta_{\mu\nu}$ with a more general background metric $G_{\mu\nu}(X)$,
where a capital letter is used to distinguish the spacetime metric from the world 
sheet metric.  From the two-dimensional point of view, this turns the action 
into that of an interacting field theory, and we must, in general, include other
counterterms as well.  To lowest order, the general action for a closed string
is
\beq
I_{\hbox{\scriptsize str}} = {1\over4\pi\alpha'}\int d^2\sigma\sqrt{g}
  \left[ g^{ab}\partial_aX^\mu\partial_bX^\nu G_{\mu\nu}(X) 
  + i\epsilon^{ab}\partial_aX^\mu\partial_bX^\nu B_{\mu\nu}(X)
  + \alpha'\,{}^{\scriptscriptstyle(2)}\!R\Phi(X) \right]
\label{g5}
\eeq
where $B_{\mu\nu}$ is an antisymmetric gauge potential for a three-form
field $H_{\mu\nu\rho}$ and $\Phi$ is a scalar, the dilaton.  Comparing with
(\ref{g1})--(\ref{g1a}), we see that the string coupling $\lambda$ is not
really an independent parameter, but is determined by the vacuum expectation 
value $\Phi_0$ of the dilaton.  If we now impose Weyl invariance---technically
by requiring that the renormalization group beta functions vanish---we find 
that $G_{\mu\nu}$, $B_{\mu\nu}$, and $\Phi$ cannot be specified arbitrarily; 
they must, rather, obey a set of equations that can be obtained from an action
\beq
I_0 = {1\over2\kappa_0^2}\int d^Dx
  \sqrt{-G}e^{-2\Phi}\left[ -{2(D-26)\over3\alpha'} +
  {}^{\scriptscriptstyle(D)}\!R[G] - {1\over12}H_{\mu\nu\rho}H^{\mu\nu\rho}
  + G^{\mu\nu} \partial_\mu\Phi\partial_\nu\Phi + O(\alpha')\right] .
\label{g6}
\eeq
This is almost the Einstein-Hilbert action for a gravitational field coupled 
to matter.  Indeed, if we rescale  the metric to ${\tilde G}_{\mu\nu} = 
e^{-4/(D-2)\,(\Phi-\Phi_0)}G_{\mu\nu}$, then the curvature term in (\ref{g6}) 
takes the standard Einstein-Hilbert form---albeit at the expense of slightly 
complicated matter couplings---with Newton's constant $G_N$ given by
\beq
8\pi G_N = \kappa_0{}^2e^{-2\Phi_0} .
\label{g6a}
\eeq
For more general supersymmetric string theories, additional background
fields appear in the action, and the rescaling of the metric gives them different
couplings to $\Phi$.  Thus string theory predicts potentially observable violations
of the equivalence principle \cite{Damour,Kaplan}.  Higher order corrections give
additional terms in the background action involving higher powers of the
curvature, but all have finite and calculable coefficients.

String theory thus contains gravity in two ways: string excitations include
``gravitons,'' and consistent propagation requires that the background 
spacetime satisfy the Einstein field equations.  The two are not independent:
if we take the background to be nearly flat, $G_{\mu\nu}=\eta_{\mu\nu}+
h_{\mu\nu}$, and expand in powers of $h$, we find that the first-order term 
in the action (\ref{g5}) is exactly the ``vertex operator'' that creates a graviton.   
  
The background field action (\ref{g6}) has a huge cosmological constant,
proportional to $D-26$.  If flat spacetime is to be even an approximate solution, 
we must require that $D=26$ (or $D=10$ for the supersymmetric extension of the 
action).  For string theory to describe our world, we must somehow hide all but four 
of these dimensions.  We can do so by ``compactifying'' spacetime, requiring that 
all but four dimensions form a small (typically Planck-scale) compact manifold, 
as in Kaluza-Klein theory \cite{quartet}. Alternatively, it could be that the 
physical processes we observe are restricted to a hypersurface in a higher 
dimensional spacetime \cite{Arkani,Randall}.  In either case, perturbative 
string theory fails to pick out a unique ``vacuum'' state; the hope, as yet unrealized, 
is that a full nonperturbative treatment might.

The need for a nonperturbative treatment goes deeper.  The string analog
of a sum of Feynman diagrams is a sum over intermediate world sheet
topologies; figure \ref{fig5}, for instance, shows a one-loop contribution
to the propagator.  String diagrams are finite order by order, but the sum
does not converge, and the series is not even Borel summable \cite{Gross}.  This 
is not quite as bad as it sounds---the same is true for Feynman diagrams in 
quantum electrodynamics, for instance---but it means that perturbative string 
theory only makes sense as an asymptotic expansion of some nonperturbative 
theory.

\subsection{Dualities and D-branes \label{secgb}}

Consider a closed bosonic string in a flat spacetime with one coordinate, 
say $X^{25}$, compactified to a circle of radius $R$.  This requires two 
changes in the mode expansion (\ref{g3}).  First, the momentum $p^{25}$ 
is no longer arbitrary, but must be quantized: $p^{25} = n/R$ ($n\in{\bf Z}$).   
Second, we must allow new ``winding modes'' $mR\sigma$ ($m\in{\bf Z}$) in 
the expansion of $X^{25}$; while these are not strictly periodic, they change 
$X^{25}$ by $2\pi mR$  as $\sigma$ varies from $0$ to $2\pi$, and are therefore 
permitted by the identification $X^{25}\sim X^{25}+2\pi R$.  It is not hard 
to compute the masses of states containing quantized momenta and winding
modes; we find
\beq
M^2 = {n^2\over R^2} + {m^2R^2\over\alpha^{\prime 2}} +
  \hbox{\it oscillator contributions} .
\label{g7}
\eeq

The mass spectrum (\ref{g7}) is invariant under the exchange
\beq
n\leftrightarrow m ,\qquad R\leftrightarrow \alpha'/R .
\label{g8}
\eeq
In fact, {\em everything\/} in the theory is invariant \cite{Nair}: string 
theory on a circle of radius $R$ is strictly equivalent to string theory on a 
circle of radius $\alpha'/R$.  A slightly more detailed analysis shows that 
the background fields in the effective action (\ref{g6}) {\em do\/} transform 
nontrivially.  From the spacetime point of view, this duality is thus a symmetry 
that relates different background fields.  This is a profound statement: it 
tells us that field configurations that are clearly distinct in general relativity 
must be treated as identical in string theory.

Next consider the limit as the compactification radius goes to zero, so the 
momentum $p^{25}$ goes to infinity.  In a field theory one would 
expect the corresponding excitations to drop out of the spectrum; the
theory would become effectively 25-dimensional.  In closed string theory,
on the other hand, $R\rightarrow0$ is equivalent to $R\rightarrow\infty$:
the winding modes become massless, and a noncompact 26th dimension  
reemerges.

Open strings, on the other hand, have no winding modes, and their  $p^{25} 
\ne0$ states really do drop out as $R\rightarrow0$.  This seems paradoxical: 
how can closed strings that see 26 dimensions interact with open strings that 
see only 25?  The answer turns out to be that the interiors of the open strings are 
still 26-dimensional, but their {\em endpoints\/} are restricted to a  collection 
of 25-dimensional hypersurfaces.  Such lower-dimensional hypersurfaces 
upon which open string endpoints are frozen are called D-branes (for
``Dirichlet membranes'') \cite{Polchinski3}.  D-branes are string theory 
solitons, and like solitons in ordinary field theory, they have masses 
inversely proportional to the coupling strength and give rise to important 
nonperturbative effects.  A recent review may be found in \cite{Johnson}.
 
The symmetry (\ref{g8}) is the simplest example of ``T-duality,''  the 
first of a great many relations among string theories that have been discovered 
in the past few years.  There are five known consistent supersymmetric string 
theories,\footnote{The bosonic string theory of section \ref{secga} is not actually 
one of them: its spectrum contains tachyons, imaginary mass excitations whose 
presence indicates an unstable vacuum.} but we now understand that they are 
all related, and related as well to 11-dimensional supergravity, through duality 
transformations.  (For a review, see \cite{Sen2}, or for  a less technical
treatment, \cite{Polchinskix}.)  In general, dualities connect one string theory 
at weak coupling with another at strong coupling, and map elementary string 
states in one theory to solitons in another.  This makes it possible to avoid some 
of the difficulties that come with strong couplings and nonperturbative effects 
by rephrasing questions as simpler but equivalent ``dual'' questions in a weakly 
coupled theory.

As more has been learned about string dualities, it has seemed more and more 
likely that string theories are limits of a larger theory, given the deliberately 
ambiguous name M theory.\footnote{``M theory'' can refer either to the 
putative theory that has 11-dimensional supergravity as its low-energy limit, 
or more broadly to the theory that includes 11-dimensional supergravity 
{\em and\/} string theories.} Perturbative string theories and 
11-dimensional supergravity live at the edges of the moduli space of M theory, 
while the interior remains unknown.  Presumably the deep questions involved in 
quantum gravity can be answered only by investigating M theory.

\subsection{The AdS/CFT correspondence}

There are currently two major programs for gaining information about M theory.
The first, ``M(atrix) theory'' \cite{BFSS}, is a quantum mechanical theory of a 
small number of $N\times N$ matrices and their supersymmetric partners, and is 
conjectured to be equivalent to M theory in a particular frame.  Remarkably, this 
simple set-up can be shown to reproduce at least linearized gravity in eleven 
dimensions, along with some nonlinear corrections.  M(atrix) theory is reviewed
in depth in reference \cite{Taylor}, and I will have no more to say about it here.  
The second, Maldacena's celebrated ``AdS/CFT correspondence'' \cite{Maldacena}, 
is a proposal that nonperturbative string theory in an asymptotically anti-de Sitter 
background is exactly dual to a flat spacetime conformal field theory in one less 
dimension.

More precisely, consider a ten-dimensional spacetime with the structure 
$M^{d+1}\times N^{9-d}$ (with $d=2$ or $4$) or an eleven-dimensional 
spacetime with the structure $M^{d+1}\times N^{10-d}$ (with $d=3$ or $6$), 
where $M$ is asymptotically anti-de Sitter and $N$ is compact.  The AdS/CFT 
conjecture is that string theory on $M\times N$ is dual to a conformal field theory 
on a flat $d$-dimensional spacetime conformal to the boundary of $M$ at spatial
infinity.  Such a correspondence requires that each field $\Phi_i$ in the string 
theory be associated with an operator ${\cal O}^i$ in the conformal field theory.  
The partition function $Z_{\hbox{\scriptsize string}}$ of the string theory depends 
on the asymptotic boundary values $\Phi_{b,i}$ of the fields---this is directly 
analogous to the boundary value dependence of the Euclidean path integral in 
section \ref{secdb}---and the proposal is that \cite{Gubser,Witten2}
\beq
Z_{\hbox{\scriptsize string}}[\Phi_{b,i}] = \left\langle
 e^{i\int_{\partial M}\Phi_{b,i}{\cal O}^i}\right\rangle_{\hbox{\scriptsize\em CFT}}
\label{g9}
\eeq
where the expectation value on the right-hand side is taken in the conformal field
theory, with the (fixed) $\Phi_{b,i}$ treated as source terms.  

Evidence for such a duality first appeared in the study of coincident D3-branes 
(that is, (3+1)-dimensional D-branes).  The string theory of $N$ such branes 
has an effective description in the small $g_cN$ limit in terms of an $\hbox{SU}(N)$ 
supersymmetric Yang-Mills theory, and in the large $g_cN$ limit in terms of  
supergravity on a spacetime with the background geometry $\hbox{AdS}_5\times S^5$.  
Since the correspondence relates weak and strong couplings, direct tests are difficult, 
but a large amount of evidence has accumulated: global symmetries, including certain 
discrete symmetries, match; spectra of chiral operators match; anomalies match; and 
phase changes at finite temperatures match.  Moreover, some correlation functions 
{\em can\/} be compared, because they are protected from large quantum corrections 
by symmetries; they, too, match.  An extensive review of these results can be found in 
\cite{Aharony}.

Given that the AdS/CFT correspondence describes gravity in ten (or eleven)
dimensions in terms of a field theory in $d<10$ dimensions, one should expect the 
relationship between degrees of freedom to be subtle.  For example, the boundary 
of anti-de Sitter space cannot directly describe the ``radial'' dependence of the 
gravitational field; that information seems to be encoded, at least in part, in the size of 
the corresponding conformal field theory excitation \cite{Susskind,Balasubramanian}.  
There are, moreover, strong arguments that localized objects in the interior of anti-de 
Sitter space must be represented by highly nonlocal operators in the conformal field 
theory \cite{Susskind2,Susskind3}.  This suggests that some of the fundamental issues 
of quantum gravity discussed in section \ref{secb}---in particular, the problem of 
locality and the reconstruction problem---are just now making their appearance 
in string theory.

\subsection{String theory and quantum gravity}

We have seen that perturbative string theory contains perturbative quantum 
gravity,  while successfully avoiding the problem of nonrenormalizability.  
Unfortunately, this does not yet mean that string theory is a quantum theory of 
gravity.  Perturbative string theory is at best an asymptotic expansion, and the 
nonperturbative theory of which it is an expansion remains largely unknown.
Nevertheless, we have some intriguing hints:
\begin{enumerate}
\item The T duality of section \ref{secgb} implies a minimum compactification
radius: string theory on a circle of radius less than $\sqrt{\alpha'}$ is completely
equivalent to string theory on a larger radius.  Similarly, investigations of high
energy string scattering suggest a modified uncertainty relation of the form
(\ref{b1}) with $L^2\sim\alpha'$ \cite{Amati}, essentially because the extra 
energy required to increase the momentum of a string probe also increases its
size.  The issue of a minimum length is delicate, since D-branes can probe
distances smaller than $\sqrt{\alpha'}$ \cite{Douglas}, but there is definitely a
suggestion of a new short-distance spacetime structure on the string (or perhaps 
the 11-dimensional Planck) scale.
\item The AdS/CFT correspondence, and in a somewhat different manner 
the M(atrix) model, provide concrete realizations of the ``holographic hypothesis'' 
\cite{tHooft2,Susskind4}.  Holography is the proposal, inspired by the behavior of 
black hole entropy, that the number of degrees of freedom of a gravitating system
in a region $R$ of space is the same as that of a system on the boundary of $R$.
(For a review, see \cite{Bigatti}.)  Such a hypothesis requires a rather drastic
reformulation of our description of fields and interactions, and presumably of
spacetime itself: it not only restricts the number of degrees of freedom, but it 
allows them to grow only much more slowly than volume.  These restrictions are 
intrinsically nonlocal, and they requires that we ultimately abandon local quantum 
field theory.  This is not an unreasonable outcome of a quantum theory of gravity,
since, as we have seen, diffeomorphism invariance already requires that observables
be nonlocal.  But it leaves us with the difficult task---not yet accomplished in 
string theory---of explaining why the Universe {\em looks\/} local.
\item As a related consequence, the AdS/CFT correspondence offers a way to
construct observables in quantum gravity, from fields of the dual conformal
field theory.  The problem is, once again, that while these observables are useful
for describing certain asymptotic properties ($S$-matrix elements, for instance),
we do not know how to recover a local spacetime description.
\item Under some circumstances, coordinates describing the locations of
D-branes become noncommuting \cite{Seiberg,Meyers}.  This suggests a possible 
role for noncommutative geometry (see section \ref{sechb}), perhaps even as a 
replacement for ordinary Riemannian geometry.
\item Even in its incomplete form, string theory suggests answers to some of the
common questions one might ask of quantum gravity.  For instance:
\begin{itemize}
\item String theory eliminates some, but not all singularities \cite{Horowitzx,%
Johnson2,Horowitz2}.  Unfortunately, we do not yet understand the physically 
important cases of cosmology and gravitational collapse.
\item String theory allows spatial topology to change 
\cite{Horowitz2,Aspinwall,Greene}.  At the same time, it restricts certain
topologies that would otherwise occur in the Euclidean path integral, possibly
eliminating the divergences in the sum over topologies described in section
\ref{secdb} \cite{Dijkgraaf}.
\item String theory provides a microscopic description of black hole entropy,
at least for some black holes.  I will discuss this further in section \ref{secib}.
\end{itemize}
\end{enumerate}

Finally, it is amusing to speculate about the connection between string theory
and loop quantum gravity.  In both theories, the fundamental excitations are 
one-dimensional objects that trace out two-dimensional world sheets/spin foams;
perhaps the two approaches are seeing different aspects of the same underlying 
structure.  There has, in fact, been some preliminary exploration of the possibility 
of  unifying string theory and loop quantum gravity \cite{Smolin4}, but the ideas 
remain highly speculative.

\section{Other approaches}\setcounter{footnote}{0}

As we have seen, attempts to quantize gravity force us to confront fundamental
questions about the nature of space and time.  In addition to the direct attacks on 
the problem that I have described above, workers in this field have tried two other 
approaches: they have looked at simpler models, and have begun to explore more 
radical ways in which the underlying assumptions of general relativity and quantum 
theory might be altered.  In this section, I will briefly describe some of these efforts.

\subsection{Simpler models \label{secha}}

Much of the difficulty in understanding quantum gravity comes from the fact
that deep conceptual issues are entangled with problems that are complicated
but ``merely technical.''  We can gain insight by looking at simpler models 
that share some of the conceptual foundations while simplifying the technical 
difficulties.  In reference \cite{Kuchar}, for example, Kucha{\v r} describes a 
number of simple physical systems, ranging from relativistic particles to coupled 
harmonic oscillators, that have been used to model pieces of the ``problem of time.''  
Systems closer to real (3+1)-dimensional general relativity include the following:

{\bf Lattices:}  By putting general relativity on a lattice, we can reduce the theory 
to a system of finitely many degrees of freedom, which may be more amenable
to both direct calculation and numerical methods.  Broadly speaking, there are two
lattice formulations of general relativity: Regge calculus \cite{Regge}, in which
spacetime is approximated as a simplicial manifold with varying edge lengths, and
the method of dynamical triangulations \cite{Ambjorn}, in which edge lengths are
fixed but the triangulation is allowed to vary.  Ashtekar variables for quantum gravity
have also been put on the lattice, and spectra of geometric operators have been
computed \cite{Loll}.  Good reviews can be found in \cite{Loll2,Ambjorn1}.

Lattice approaches to quantum gravity typically focus on the Euclidean path integral 
of section \ref{secdb}, which can be approximated as a finite sum over lattice 
configurations and can be evaluated, for example, by Monte Carlo simulations.  The 
results indicate the existence of two phases: for large values of Newton's constant 
$G$ a ``crumpled'' phase with small curvature, high Hausdorff dimension, and high 
connectivity, and for small $G$ a tree-like ``branched polymer'' phase with large 
curvature and Hausdorff dimension 2.  Neither looks much like a classical spacetime, 
and there is now good evidence that the phase transition between them is first order, 
so no new continuum physics is expected at the transition.   But some exciting 
new results \cite{Ambjorn1,Ambjorn2} indicate that lattice models for the 
{\em Lorentzian\/} path integral behave very differently, with a semiclassical limit 
that looks much more like standard general relativity.  It is too early to assess these 
results, but they suggest that a careful treatment of  causality may be more important 
than has been generally appreciated. 

{\bf Mini- and midisuperspaces}: A more drastic simplification of general relativity 
can be obtained by ``freezing out'' all but a few of the degrees of freedom of the metric 
\cite{DeWitt7,Misner2,Ryan,Wiltshire}.  By choosing an ansatz in which the metric 
depends only on a few parameters, we can reduce the space of metrics---the superspace 
of section \ref{secea}---to a finite-dimensional ``minisuperspace,''  such as the space 
of homogeneous cosmologies.  Alternatively, we may keep enough degrees of freedom 
to allow an infinite-dimensional ``midisuperspace'' \cite{Kuchar6}, such as the space 
of cylindrical gravitational waves, while still simplifying enough to allow quantization.  
Minisuperspace models have dominated work in quantum cosmology, and mini- 
and midisuperspaces have been used to investigate the Wheeler-DeWitt equation,
Lorentzian and Euclidean path integrals, and reduced phase space quantization.  Such 
models have been testing grounds for the notion of ``evolving constants of motion'' 
\cite{Tate} and for the reconstruction of geometric quantities from diffeomorphism-%
invariant quantum observables \cite{Ashtekar10}, and they have provided interesting 
insights into the possibilities of quantum fluctuations of the light cone \cite{Ashtekar10}
and certain unexpectedly large quantum gravitational effects \cite{Angulo}.

Caution must be used in interpreting such calculations, however.  The minisuperspace
approximation can be tested by embedding a model with high symmetry and few 
variables into one with lower symmetry and more variables, in effect ``unfreezing'' 
some degrees of freedom.  The behavior can turn out to be qualitatively very different 
\cite{Kuchar7}: minisuperspace models can miss important physics.

{\bf Lower dimensional models:} Classical general relativity becomes much simpler 
in fewer than four spacetime dimensions, and this simplification carries over to the 
quantum theory as well.  In particular, general relativity in 2+1 dimensions has no 
field degrees of freedom: in a canonical formulation, we have six degrees of freedom 
in $\{q_{ij},\pi^{ij}\}$ minus three constraints and three symmetries.  This does 
not mean that the theory is trivial---in the presence of point particles or nontrivial 
topologies, a finite number of global degrees of freedom remain---but it means that 
quantum field theory reduces to much more manageable quantum mechanics. The 
(2+1)-dimensional model has no Newtonian limit, and is not physically realistic; 
but as a diffeomorphism-invariant theory of spacetime geometry, it shares the basic 
conceptual underpinnings of (3+1)-dimensional general relativity, and provides a 
valuable test bed for quantization.

Many of the quantization programs described here---phase space path integrals, 
reduced phase space quantization, covariant canonical quantization, Dirac 
quantization with first-order variables, loop quantum gravity, and various exact 
lattice methods---can be carried out in full in 2+1 dimensions, at least for simple 
spatial topologies.  (For a review, see \cite{Carlip2}.)  Others, notably the Euclidean 
path integral and the Wheeler-DeWitt equation, remain difficult, but can be explored 
in more detail.  Results from 2+1 dimensions provide an ``existence proof'' for 
quantum gravity, showing that general relativity can be quantized, at least in a simple 
setting, without any need for additional ingredients.  In particular, the reconstruction 
problem and the problems of time can be resolved, and a sensible classical limit 
can be found.  The model also provides a ``nonuniqueness proof'': different consistent 
approaches can lead to quantum theories that differ in rather fundamental ways, for 
instance in their answer to the question of whether spacetime has a minimum length.  

In two spacetime dimensions, the Einstein-Hilbert action is a topological invariant, 
and the Einstein tensor is identically zero.  One can write a more general action,  
though:
\beq
I = \int d^2x\sqrt{-g}\left( 
    a\varphi R + bg^{\mu\nu}\partial_\mu\varphi\partial_\nu\varphi + V[\varphi]\right) ,
\label{h1}
\eeq
where $\varphi$ is a scalar, the dilaton, $a$ and $b$ are constants, and $V$ is an 
arbitrary potential.  The special case $V\sim e^{\lambda\varphi}$ is known as 
Liouville theory; it arises frequently in string theory, and has been studied extensively 
\cite{Seiberg2}, although some key questions remain.  Dilaton gravity coupled to $N$ 
scalar fields and evaluated in the $1/N$ expansion has proven useful in studying 
the quantum mechanics of black hole formulation and evaporation \cite{CGHS}, and 
a variety of approaches to the quantization of dilaton models have been explored 
\cite{Cangemi,Benedict,Martinez,Kuchar8}.

{\bf Planckian scattering:} The study of scattering at Planckian energies 
would seem an unlikely place to try to gain control over quantum gravity.  But 
scattering involves two scales, center of mass energy and momentum transfer, and 
a new approximation---essentially the eikonal approximation---is available for 
small momentum transfer \cite{Verlinde,Kabat}.  The quantum dynamics of the 
longitudinal gravitational modes turns out to be described by a topological field 
theory with $S$-matrix elements that have some resemblance to string amplitudes, 
and there is an intriguing appearance of noncommuting ``ingoing'' and ``outgoing'' 
coordinates, suggesting a new uncertainty relationship that might be relevant for 
black hole evaporation \cite{tHooft3}.

\subsection{Radical departures \label{sechb}}

It is safe to say that most people working in quantum gravity expect that the theory 
will eventually lead to radical changes in our understanding of space and time.  A few 
are more ambitious: they argue that radical changes in our starting point may be a 
precondition for quantizing gravity.  It may be necessary, for example, to reformulate 
classical geometry in a way that makes causal relations more fundamental, or to 
somehow ``quantize'' spacetime points, or to even do away with the idea of points 
altogether.  Work along these lines has not yet led to any physical breakthroughs, 
but perhaps that is too much to ask, given that more conventional approaches have 
not been terribly successful either.  A sampling of the more radical new proposals 
is the following:

{\bf Causal sets:}  A Lorentzian metric determines both a geometry and a causal 
structure, that is, a partial ordering that specifies which events lie to the future of 
any given event.  While the causal structure seems much weaker, it actually determines 
the metric up to a conformal factor.  The causal set program \cite{Sorkin} takes the 
causal structure as primary, and starts with a finite set of points with a causal ordering; 
the conformal factor is then (approximately) recovered by counting points.  Computer 
simulations of simple models have shown evidence for a continuum limit \cite{Rideout}, 
and the causal set proposal has recently been combined with the loop representation 
to formulate ``causal spin foams'' \cite{Markopoulou2}.

{\bf Twistors:}  The twistor program \cite{Penrose2} also gives primacy to the 
causal structure of spacetime, essentially by viewing spacetime as a collection of 
null geodesics.  Points are now derived quantities---a point is represented by a 
sphere in four-complex-dimensional twistor space,  corresponding physically to 
the celestial sphere at that point.  The goal is to translate all spacetime physics into 
the ``more primitive''realm of twistor space.  The program has had success in treating 
massless fields in flat spacetime, but gravity has proven more elusive; for a summary 
of some recent developments, see \cite{Penrose4}.

{\bf Null surface formulation}:  The null surface formalism \cite{Kozameh} also 
takes null geodesics to be fundamental, and rewrites general relativity as a theory of 
null surfaces.  At each point, a function $Z(x,\zeta,{\bar\zeta})$ gives a two-sphere's
worth of surfaces; when a single metricity condition is imposed, these become null
surfaces, with $(\zeta,{\bar\zeta})$ acting as coordinates on the celestial sphere.
The Einstein field equations can be rewritten in terms of $Z$, and the free data that
determines $Z$ can, at least in principle, be quantized at null infinity.  The resulting 
picture of spacetime has ``fuzzy points'' and ``fuzzy null cones.'' 

{\bf Noncommutative geometry:} Given that quantizing gravity means ``quantizing 
the structure of spacetime,'' and that quantum mechanics is characterized by the
existence of noncommuting observables, it is natural to look for a way to make
spacetime itself noncommutative.  Heuristically, a noncommutative geometry is 
simply one in which spacetime coordinates fail to commute.  This idea dates back
to Snyder \cite{Snyder}, who suggested in 1947 that quantum field theory on a 
spacetime with noncommutative coordinates might be less divergent while
still preserving Lorentz invariance.  A recent resurgence of interest has been 
inspired in part by the pioneering mathematical work of Connes \cite{Connes}, 
and in part by the discovery that certain D-brane configurations in string theory 
naturally involve noncommuting coordinates \cite{Seiberg,Meyers}.

The notion of noncommutative geometry is a bit tricky, especially in general
relativity, where coordinates---and, indeed, points---have no independent
physical meaning.  To make the idea sufficiently general and well defined \cite{%
Connes,Madore}, one must take a detour first suggested by Geroch \cite{Geroch2}, 
and reexpress the spacetime manifold $M$ in terms of the algebra of functions on 
$M$.  Much is now known about how to obtain the ordinary picture of spacetime 
from such an algebra of functions; the geometry, for example, can be reconstructed 
from the eigenvalues of the Dirac operator $\slash D$.  Geometry can then be made 
noncommutative by appropriately generalizing to a noncommutative algebra of 
functions.

Fields on a noncommutative geometry are themselves noncommuting, even at
the classical level.  In some cases, though, a field theory in a noncommutative 
geometry is equivalent to a theory involving ordinary commuting fields, but with 
products of fields in the action replaced by nonlocal ``Moyal'' \cite{Moyal} or 
$\star$-products.\footnote{For a recent review of quantum field theory on
noncommutative spacetime, see \cite{Douglas2}.}    The resulting quantum 
theories are nonlocal, and exhibit some peculiar features, including the coupling of 
high- and low-energy degrees of freedom (``UV/IR mixing'') \cite{Minwalla,Vaidya}.  
They are sometimes unitary, but need not be \cite{Gomis}.  In general relativity, 
the $\star$-product approach requires the introduction of complex metrics; there 
have been some interesting preliminary investigations \cite{Chamseddine,Moffat},  
but it is not yet clear where they will lead.  A somewhat different approach, based 
more directly on Connes' general formulation of noncommutative geometry, can 
give rise to a classical model that seems to naturally incorporate both gravity and 
the Standard Model of high energy physics \cite{Connes2}.

For the most part, noncommutative geometry is still ``classical''---it does not involve 
Planck's constant---and much remains unknown about how to quantize the resulting 
theories.  The program has had an interesting impact on other approaches to quantum
gravity, though: it has led to the realization that the eigenvalues of $\slash D$ provide 
a nice set of nonlocal, diffeomorphism-invariant observables, whose classical Poisson 
brackets can be computed \cite{Landi}.  At least for metrics with Riemannian signature, 
these provide a nearly complete characterization of the classical geometry, thus giving 
us the first good candidates for a (nearly) complete set of diffeomorphism-invariant 
observables.

{\bf Other possibilities:} One radical option is to quantize not just the geometry, 
but the point set topology of spacetime \cite{Ishamx}.  Another is to change the
foundations of axiomatic quantum field theory to make them compatible with general 
covariance \cite{Fredenhagen}, or, more drastically, to make quantum mechanics 
itself nonlinear and allow gravity to cause wave function collapse \cite{Penrose}.  
Or perhaps we need to begin with an even more fundamental ``pregeometry'': 
't Hooft has suggested that a deterministic but dissipative system should underlie
quantum mechanics at the Planck scale \cite{tHooftx}, and Wheeler has even 
proposed the calculus of propositions as a starting point\cite{MTW}!

\section{How will we know we're right?}

We finally turn to the most daunting problem of quantum gravity, the nearly complete
lack of observational and experimental evidence that could point us in the right direction
or provide tests for our models.  The ultimate measure of any theory is its agreement with 
Nature; if we do not have any such tests, how will we know whether we're right?

It may be that there is only one self-consistent quantum theory of gravity, and that 
any answer is equivalent to any other.  But progress in (2+1)-dimensional quantum 
gravity casts doubt on this possibility: in that setting, at least, there are consistent but
physically inequivalent quantizations \cite{Carlip2}.  Similarly, there is now strong 
evidence for dramatic differences between lattice models for Euclidean and Lorentzian
path integrals \cite{Ambjorn2}.

Fortunately, though, the picture is not so grim.  There are a number of places at which 
quantum gravity can already come into contact with observation, and more may 
appear in the not-too-distant future.

\subsection{The classical limit \label{secia}}
  
The ``zeroth test'' of any quantum theory of gravity is its ability to reproduce the
successes of classical general relativity.  This is not trivial: finding the classical limit 
requires that we solve the ``reconstruction problem,'' the problem of recovering
local geometry from nonlocal observables.   The problem is not just technical---it
is by no means automatic that a theory obtained by ``quantizing general relativity'' 
will have general relativity as its classical limit.   As we saw in section \ref{secfb}, 
for example, there are strong arguments that the best known definition of the 
Hamiltonian constraint in loop quantum gravity leads to a theory whose classical
limit is not general relativity.  Similarly, lattice models for the Euclidean path
integral fail to predict smooth four-geometries, and naive perturbative string
theory predicts a 10-dimensional universe.  

In particular, a successful quantum theory of gravity must predict, or at least
allow, a classical limit with a nearly vanishing cosmological constant.  This 
is a major challenge, since straightforward effective field theory arguments 
predict that $\Lambda$ should receive quantum corrections of order 
$L_{\hbox{\scriptsize Planck}}^{-2}$, some 120 orders of magnitude
larger than the observational limit \cite{Weinberg3,Carroll}.  It is possible 
that there is a ``generic'' solution to this problem---there are tentative 
arguments, described in section \ref{secdb}, that fluctuations in spacetime 
topology could suppress positive \cite{Coleman} or negative \cite{Carlip4} 
values of $\Lambda$---but a successful quantization of gravity should presumably 
provide a detailed mechanism.  The demonstration of such a mechanism would 
be a strong argument for the validity of an approach to quantization.

Arguably, quantum gravity should provide even more.  Our Universe is not a
generic solution of the field equations, but a particular one, and quantum
gravity may have something to say about which one.  For example:
\begin{itemize} 
\item We do not know the large scale topology of the Universe, but it may
soon be measurable \cite{Cornish}. If quantum gravity permits fluctuating
spatial topology, it should be possible to obtain at least a probabilistic
prediction of this topology.
\item In inflationary cosmologies, the large scale structure of the Universe is 
seeded by early quantum fluctuations, whose spectra determine the cosmic 
microwave background anisotropies and the structure observed today \cite{Liddle}.  
But the details of these spectra depend on assumptions about the initial state 
\cite{Lesgourgues}, which may in turn be affected by quantum gravity.  Moreover, 
since inflation can stretch out initial smaller-than-Planck-size fluctuations to 
observable scales, predictions may be sensitive to details of Planck length physics,
although a careful analysis shows considerably less sensitivity than one might
naively expect \cite{Brandenberger,Niemeyer,Mersini,Tanaka,Starobinsky,Easther}.
\item Some proposals---notably the Hartle-Hawking \cite{Hartle} and the 
Vilenkin \cite{Vilenkin} boundary conditions---make specific predictions about 
the quantum state of the Universe.  String theory may do more, since the string
vacuum determines the spectrum and gauge group of elementary particles.  One 
may argue about whether such predictions are part of quantum gravity proper
or whether they are added assumptions, but they at least require a well-formulated 
quantum theory of gravity as a setting.
\end{itemize}

\subsection{Black hole thermodynamics \label{secib}}

Even in the absence of a quantum theory of gravity, we have one robust prediction
of such a theory: the existence and spectrum of black hole Hawking radiation.
Hawking radiation is a ``semiclassical'' prediction,  originally discovered in the
study of quantum field theory on a fixed curved background \cite{Hawking5}, and
subsequently confirmed in a remarkable number of different computations,
ranging from the saddle point approximation of the Euclidean path integral 
\cite{Gibbons3} to the investigation of symplectic structure of the space of 
solutions \cite{Wald4} to the calculation of amplitudes for black hole pair 
production \cite{Garfinkle,Dowker}.  Any theory that fails to reproduce this 
prediction is almost surely wrong.

Black hole radiation implies a temperature and an entropy, the Bekenstein-Hawking
entropy \cite{Hawking5,Bekenstein}
\beq
S_{\hbox{\scriptsize BH}} = {A_{\hbox{\scriptsize horizon}}\over4\hbar G} .
\label{i1}
\eeq
In ordinary thermodynamics, entropy is a measure of the number of states, and a
successful quantum theory of gravity ought to allow us to describe and enumerate
the states responsible for $S_{\hbox{\scriptsize BH}}$.  Both string theory and
loop quantum gravity have been partially successful in meeting this challenge.
String theory reproduces (\ref{i1}) for a large class of extremal and near-extremal
black holes by counting states at weak coupling; the number of states is then
protected by supersymmetry as the coupling increases.  For nonextremal black
holes, string theory continues to predict an entropy proportional to horizon
area, but the constant of proportionality becomes hard to calculate \cite{Peet}.
Loop quantum gravity leads to an entropy for both extremal and nonextremal
black holes of the form (\ref{i1}), but with a proportionality constant that
depends on the Immirzi parameter \cite{Ashtekar11,Ashtekar12}.  The peculiar 
choice $\gamma = \ln 2/\pi\sqrt{3}$ seems to be necessary to reproduce the 
Bekenstein-Hawking result, but the same choice then holds for a wide variety
of different black holes.

It has recently been suggested that the value of the Bekenstein-Hawking entropy 
may be fixed ``universally'' by a broken conformal symmetry at or near the horizon, 
which is inherited in turn from the diffeomorphism invariance of general
relativity \cite{Carlip9}.  If this is true, then a quantum prediction of the entropy 
is just a test that the group of diffeomorphisms is represented correctly.  Even so, 
a true quantum theory of gravity will not only predict the entropy (\ref{i1}); it
will provide a concrete realization of the relevant degrees of freedom.  Different
theories may also differ in their one-loop corrections to the entropy, although
there are some interesting signs of ``universality'' there as well \cite{Carlip10}.

\subsection{Direct tests}

The characteristic scale for quantum gravity is the Planck energy, 
$E_{\hbox{\scriptsize Planck}}\sim 10^{19}$ GeV.  This is so far out of the range 
of experiment that direct observational tests have long seemed impossible.  In the 
past few years, though, a number of tests have been proposed, based on two ideas: 
that we can detect very small deviations from otherwise exact symmetries, and 
that we can integrate over long distances or times to observe very small collective 
effects.  For the most part, these proposals remain highly speculative, but they
are at least somewhat plausible.  A recent review of many of these ideas may be 
found in \cite{Amelino}.  In particular:
\begin{itemize}
\item Quantum gravity may lead to violations of the equivalence principle, either
generically for superpositions of mass states \cite{Adunas} or in specific models,
for instance from dilaton couplings in string theory \cite{Damour,Kaplan}.  These 
effects may be detectable in future precision tests of the equivalence principle and in 
atomic and neutron interferometry.
\item Quantum gravity may lead to violations of CPT invariance, for instance
through the formation of virtual black holes \cite{Ellis,Kostelecky}.  Present 
experimental limits are approaching the level at which such effects might be 
observable \cite{CPLEAR}.  Gravitational effects may also lead to violations
of other global symmetries such as CP, with observable consequences that may
be quite sensitive to the Planck scale structure of spacetime \cite{Kallosh}.
\item Quantum gravity may distort the dispersion relations for light and 
neutrinos over long distances, leading to a frequency-dependent speed of light 
\cite{Amelino2,Gambini3,Alfaro}.  This effect is potentially testable through the 
observation of gamma ray bursts; current limits are near the scale at which quantum 
gravity might be significant \cite{Biller}.  The effect may also depend on polarization,
and tests of gravity-induced birefringence may be within observational reach
\cite{Gleiser}.
\item A few physicists have suggested that the interferometers being built for 
gravitational wave detection are so sensitive that they could potentially observe 
quantum fluctuations in the geometry of space \cite{Amelino3,Ng}.  Data from 
the existing 40-meter interferometer at Caltech can already be used to rule 
out one simple guess, that the arms of the interferometer undergo random 
Planck length fluctuations at a rate of one per Planck time.  Different models 
predict very different spectra of length fluctuations, but some of these may be 
testable in future interferometers, although such claims remain quite controversial.
\item Quantum gravitational effects near the Planck mass affect renormalization
group flows and low energy coupling constants in grand unified theories \cite{Hall}.
Unfortunately, this ``Planck smearing'' mainly makes it hard to test GUTs rather
than making it easy to test quantum gravity.
\item It has been proposed that the use of intense lasers to accelerate electrons may
make it possible to (indirectly) observe Unruh radiation, the counterpart of 
Hawking radiation for an accelerating particle in flat spacetime \cite{Chen}.  One 
may argue about whether this is a direct test of quantum gravity, but it is certainly 
at least a test of quantum field theoretical predictions of the sort that go into 
quantum gravity.
\item Another indirect test may come from condensed matter analogs of black holes, 
which should emit ``Hawking radiation'' phonons from sonic horizons, regions
at which the fluid flow reaches the speed of sound \cite{Unruhx,Visser}.  Tests
may be possible in the not-too-distant future in Bose-Einstein condensates
\cite{Garay2}, superfluid helium 3 \cite{Volovik}, and ``slow light'' in dielectrics
\cite{Piwnicki}.
\end{itemize}

These proposed experiments do not, for the most part, test specific models of
quantum gravity, largely because the available models cannot yet make clear
enough predictions.  They do, however, test Planck scale physics of the sort that
should generically be affected by quantum gravity.  Although it is not at all certain
that these tests are actually feasible, the possibility of directly observing quantum
gravitational effects is beginning to be taken seriously.

\subsection{Large extra dimensions and TeV-scale gravity}

In the past few months, there has been a burst of interest in possible experimental
tests for a particular class of models of quantum gravity.  These ``TeV scale gravity''
or ``brane world'' models \cite{Arkani,Randall} postulate additional ``large'' 
dimensions beyond the four we observe.  The idea of extra dimensions is certainly 
not new, and as we saw in section \ref{secga}, string theory requires them.  But 
while normal Kaluza-Klein theories hide the extra dimensions by compactifying 
them to sizes on the order of the Planck length, TeV-scale gravity allows dimensions 
as large as a millimeter.  

In these models, we do not observe the extra dimensions, not because they are too 
small to see, but because the fundamental particles and interactions we can see are 
confined to a four-dimensional ``brane.''  The idea of such confinement is natural 
in string theory, where  fields associated with the ends of open strings can be stuck 
on D-branes, but it occurs in other contexts as well: fields can be trapped on 
topological defects or gravitationally bound to a membrane.\footnote{See the 
note at the end of \cite{Visser2} for some historical references.}  One of the 
chief attractions of such models is that they offer a simple answer to the question 
of why the Planck energy is so high compared to the energy scales of the other
fundamental interactions.  The answer is that it is not: the four-dimensional
Planck scale we observe is either
\beq
{1\over 8\pi G_4} \sim {V_n\over 8\pi G_{4+n}}
\label{i2}
\eeq
for $n$ compact dimensions of volume $V_n$, or 
\beq
{1\over 8\pi G_4} \sim {1\over 8\pi G_{n+4}}
  \int_0^{y_{\mathit max}}d^ny_\perp\, e^{-2k|y_\perp|}
\label{i3}
\eeq
for $n$ noncompact dimensions with coordinates $y_\perp$ and an appropriate
``warped'' metric.  By adjusting the number and size of the extra dimensions, it is
relatively easy to obtain a theory in which the fundamental (4+n)-dimensional
Planck scale is a few TeV, on the order of the electroweak scale.

TeV-scale gravity models are no easier to quantize than ordinary four-dimensional
general relativity, and all of the fundamental conceptual problems remain.  But 
even as low energy effective theories, they generate a host of testable predictions, 
ranging from a breakdown of the inverse square law for gravity at short distances 
to missing energy carried off into the extra dimensions to the emergence of towers 
of Kaluza-Klein particles at accelerator energies.  At this writing, there is not yet 
a good review article on this subject, though several are in preparation; reference
\cite{Giudice} gives a good sample of the work being done on the phenomenology
of these models, while reference \cite{Shiromizu} gives a careful general relativistic
treatment of induced gravity on the brane.

\subsection{Other desiderata}

There are a number of other desirable features we might hope to find in a 
quantum theory of gravity, which, while not directly testable, could serve as signs 
that we are on the right track.  Notable among these are the treatment of singularities 
in general relativity and divergences in quantum field theory, and the ability to resolve 
the ``black hole information paradox.''

Wheeler has long argued that the gravitational collapse reveals a fundamental
breakdown of general relativity, which must be fixed by quantum gravity 
\cite{MTW}.  This is a delicate issue:  singularities are already difficult to define 
in classical general relativity, and it is not at all clear how to extend the classical 
definitions to quantum gravity (the ``reconstruction problem'' again).
Still, we can at least demand that the theory exclude such phenomena as infinite 
densities and wave functions that disappear off ``edges'' of spacetime.  

Quantum gravity should probably not treat all singularities the same way, though.
Some classical singularities---notably those that occur in gravitational collapse and 
big bang cosmology---are a consequence of normal physical processes, and one might 
reasonably expect quantum gravity to ``smooth them out'' into nonsingular states.  But 
other classical singularities should not be smoothed out; they must rather be forbidden 
from the start \cite{Horowitzx}.  In particular, the positive energy theorem of classical 
general relativity holds only if certain singular configurations such as negative mass 
Schwarzschild solutions are excluded.  A quantum theory that ``smoothed out'' the 
singularity of the negative mass Schwarzschild metric would then admit negative mass
states, and would have no stable vacuum.  String theory provides an interesting example
of the excision of a particular naked singularity, the ``repulson'': it turns out that effects
related to string duality prevent the singularity from ever forming in the first place
\cite{Johnson2}.

On the flip side, there is an old hope that quantum gravity will eliminate the
divergences of quantum field theory by providing a natural Planck scale cutoff 
\cite{Landau,Pauli,Deser}.  There are a number of arguments that make this
suggestion plausible.  Divergences come from the contributions of very high 
energy virtual particles, and at high energies quantum gravity cannot be 
ignored.  Gravitational fluctuations could ``smear out'' light cones and the 
associated infinities; the minimum length implied by the generalized uncertainty 
relations (\ref{b1}) could dramatically affect short distance behavior; the negative 
contribution of gravitational self-energy could cancel the divergent self-energy 
of quantum field theory, as it does in classical electromagnetism; and high energy 
virtual particles might collapse into black holes, which would at the very least 
make existing computations of divergences unreliable.  

None of these arguments is decisive, and such an ultraviolet cutoff is certainly not 
a prerequisite for a successful quantization, but a theory that provided such a cutoff 
would surely be attractive.  Ideally, a finite cutoff might even allow us to predict
some fundamental quantities like particle masses.  String theory, of course,
is even more ambitious: the hope is that a unique nonperturbative string 
vacuum would fix the entire particle spectrum and gauge group of the
Standard Model.  With or without new Planck scale observations, such a 
successful ``retrodiction'' of the Standard Model would be a decisive test of 
the theory.

Finally, the little we know about semiclassical gravity has led to a paradox
that the full quantum theory will have to resolve.  Consider a process in which
matter in a pure quantum state collapses to form a black hole, which subsequently 
evaporates via Hawking radiation.  If, as semiclassical calculations indicate,
Hawking radiation is truly thermal, then this process represents evolution from
a pure initial state to a mixed final state.  Entropy has increased, and ``information''
has been lost.  But standard quantum mechanics forbids such evolution: at the
microscopic level, quantum mechanical entropy is conserved, and a pure state
can evolve only into a pure state.  The literature is filled with ideas for resolving 
this ``black hole information paradox''  \cite{Giddings2}:  Hawking radiation 
may have complex hidden correlations; black hole evaporation may end with 
information-rich ``remnants''; or perhaps quantum mechanics should be altered
to allow pure states to become mixtures, or to introduce a ``complementarity''
between states inside and outside a horizon.   But none of the suggestions is yet very 
convincing; it may well take a full quantum theory of gravity to give a satisfactory 
explanation.

\section{Where we stand}

In classical general relativity, the ``force'' we call gravity is a consequence of the
geometry of spacetime.  Quantizing gravity means, conservatively, quantizing
the structure of space and time; or, more radically, eliminating space and time 
altogether as fundamental attributes of the Universe and replacing them with 
something new.  This is an ambitious goal, and it should not be so surprising that 
we have not yet succeeded.

Nevertheless, we have learned a great deal in the past 70 years of research.  We 
now know a quite a lot about what doesn't work: we cannot, for instance, simply
treat general relativity as an ordinary quantum field theory.  We also know much
more about what is needed for success: we must, for instance, understand how to
approximately reconstruct a classical spacetime from nonlocal diffeomorphism-%
invariant observables.

The two main contemporary programs of quantum gravity, string theory and loop
quantum gravity, are just now beginning to confront the fundamental issues.  This 
should not be interpreted as a criticism.  One cannot solve the problem of time or
the spacetime reconstruction problem by simply thinking hard about time and
observables;  it has taken tremendous work to clear away enough of the underbrush 
to even pose the questions as physics rather than philosophy.  The crucial tests,
though, are still to come.  Can string theory find good nonperturbative
observables that describe local physics, and explain how to connect these to the
gravitational predictions of perturbative string theory?  Can loop quantum gravity
solve the Hamiltonian constraint, and reconstruct good semiclassical states from
the solutions?  Can any of the more radical approaches move into the realm of
predictive theories?

The next decade promises to  be an interesting one.

\vspace{1.5ex}
\begin{flushleft}
\large\bf Acknowledgments
\end{flushleft}

I would like to thank John Baez, Gary Horowitz, Peter Salzman, and Sachindeo
Vaidya for catching errors in earlier drafts of this review.  I'm afraid I didn't give 
them time to find {\em all\/} the errors; those that remain are mine.  This work 
was supported in part by U.S.\ Department of Energy grant DE-FG03-91ER40674.

\end{document}